\documentclass[longauth,usenatbib]{aa}
\usepackage{graphicx}
\usepackage{txfonts}
\usepackage{lscape}
\usepackage{amssymb,amsmath}
\usepackage{rotating}
\usepackage{xcolor}

\def\msun{M_\odot}
\def\fs{f_{\mathrm S}}

\def\I0st{{I_{\mathrm 0}^{\mathrm{st}}}}
\def\V0{V_{\mathrm 0}}

\def\tE{t_{\mathrm E}}

\def\te{t_{\mathrm E}}
\def\t0{t_{\mathrm 0}}

\def\u0{u_{\mathrm 0}}

\def\piE{\pi_{\mathrm{E}}}
\def\piEvec{\overrightarrow{\pi_{E}}}
\def\piEE{\pi_{\mathrm{EE}}}
\def\piEN{\pi_{\mathrm{EN}}}
\def\murel{\mu_{\mathrm{rel}}}

\def\thetaE{\theta_{\mathrm{E}}}
\def\pirel{\pi_{\mathrm{rel}}}
\def\piS{\pi_\mathrm{S}}
\def\piL{\pi_\mathrm{L}}
\def\DL{D_{\mathrm{L}}}
\def\DS{D_{\mathrm{S}}}
\def\ML{M_{\mathrm{L}}}

\newcommand{\ie}{{i.e.},}

\newcommand{\gaia}{{\it{Gaia}}\xspace}
\newcommand{\Gaia}{{\it{Gaia}}\xspace}

\usepackage[breaklinks=true]{hyperref}
\newcommand{\orcit}[1]{\protect\href{https://orcid.org/#1}{\protect\includegraphics[width=8pt]{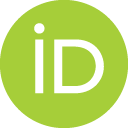}}}

\begin{document} 

  \title{Lens mass estimate in the Galactic disk extreme parallax microlensing event Gaia19dke}

\titlerunning{Lens mass estimate in Gaia19dke}
\authorrunning{M. Maskoliunas et al.}

   \author{
   M. Maskoli\={u}nas\orcit{0000-0003-3432-2393}\inst{1},
   {\L}. Wyrzykowski\orcit{0000-0002-9658-6151}\inst{2},
   K. Howil\inst{2}, 
   K. A. Rybicki\inst{44,2}, 
   P. Zieli\'{n}ski\orcit{0000-0001-6434-9429}
   \inst{3}, 
   Z. Kaczmarek\inst{4}, 
   K. Kruszy\'{n}ska\inst{2}, 
     M. Jab{\l}o{\'n}ska\inst{2},   
   J. Zdanavi\v{c}ius\inst{1},
   E. Pak\v{s}tien\.{e}\orcit{0000-0002-3326-2918}\inst{1},
   V. \v{C}epas\inst{1},
     P. J. Miko{\l}ajczyk\orcit{0000-0001-8916-8050}\inst{2,8}, 
  R. Janulis\inst{1},
    M. Gromadzki\inst{2}, 
  N. Ihanec\inst{2}, 
  R. Adomavi\v{c}ien\.{e}\inst{1},
  K. \v{S}i\v{s}kauskait\.{e}\inst{1},
  M. Bronikowski\inst{2,7},    
  P. Sivak\inst{2}, 
  A. Stankevi\v{c}i\={u}t\.{e}\inst{2}, 
  M. Sitek\inst{2}, 
  M. Ratajczak\inst{2}, 
  U. Pylypenko\inst{2}, 
    I. Gezer\inst{5}, 
  S. Awiphan\inst{9},        
  E. Bachelet\inst{10},      
  K. B\k{a}kowska\inst{3}, 
  R. P. Boyle\inst{12},         
  V. Bozza\inst{32,33}, 
  S. M. Brincat\orcit{0000-0002-9205-5329}\inst{13},       
  U. Burgaz\orcit{0000-0003-0126-3999}\inst{11},        
  T. Butterley\inst{29}, 
  J.~M.~Carrasco\orcit{0000-0002-3029-5853}\inst{14,48,49},      
  A. Cassan\inst{38}, 
  F. Cusano\inst{15},        
  G. Damljanovic\orcit{000-0002-6710-6868}\inst{6}, 
  J. W. Davidson\inst{46}, 
  V. S. Dhillon\inst{22},  
  M. Dominik\orcit{0000-0002-3202-0343}\inst{39},  
  F. Dubois\inst{16},        
  H. H. Esenoglu\orcit{0000-0003-3531-7510}\inst{17},      
  R. Figuera Jaimes\inst{34,\ref{inst:Pontificia}},         
  A. Fukui\orcit{0000-0002-4909-5763}\inst{19},         
  C. Galdies\orcit{0000-0002-8908-0785}\inst{20},       
  A. Garofalo\inst{15},   
  V. Godunova\orcit{0000-0001-7668-7994}\inst{21},      
  T. G\"uver\orcit{0000-0002-3029-5853}\inst{17,18},         
  J. Heidt\inst{22},         
  M. Hundertmark\orcit{0000-0003-0961-5231}\inst{36}, 
  I. Izviekova\inst{3}, 
  B. Joachimczyk\inst{3}, 
  M.K. Kami{\'n}ska\orcit{0000-0001-8049-196X}\inst{43},
  K. Kami{\'n}ski\orcit{0000-0002-5688-7304}\inst{43},
  S. Kaptan\orcit{0000-0003-0873-0983}\inst{17,18}, 
  T. Kvernadze\inst{24},     
  O. Kvaratskhelia\inst{24},     
  S. Littlefair\inst{22},  
  O. Michniewicz\inst{24},   
  N. Nakhatutai\inst{35},        
  W. Og{\l}oza\orcit{0000-0002-6293-9940}\inst{42},  
  R. Ohsawa\orcit{0000-0001-5797-6010}\inst{45},     
  J. M. Olszewska\orcit{0000-0003-3579-4253}\inst{43}, 
  M. Poli{\'n}ska\orcit{0000-0003-4473-3372}\inst{43}, 
  A. Popowicz\inst{25},      
  J. K. T. Qvam\inst{51},
  M. Radziwonowicz\inst{2}, 
  D. E. Reichart\inst{47}, 
  A. S{\l}owikowska\orcit{0000-0003-4525-3178}\inst{37, 3}, 
  A. Simon\orcit{0000-0003-0404-5559}\inst{30, 31},         
  E. Sonbas\orcit{0000-0002-6909-192X}\inst{40,41} 
  M. Stojanovic\orcit{0000-0002-4105-7113}\inst{6}
  Y. Tsapras\orcit{0000-0001-8411-351X}\inst{36},
  S. Vanaverbeke\inst{16},   
  J. Wambsganss\inst{36},
  R. W. Wilson\inst{29}, 
  M. {\.Z}ejmo\inst{24},         
  S. Zola\inst{28},          
  }

\institute{
Institute of Theoretical Physics and Astronomy, Vilnius University, Saulėtekio Av. 3, 10257 Vilnius, Lithuania, \email{marius.maskoliunas@tfai.vu.lt}
\and
Astronomical Observatory, University of Warsaw, Al.~Ujazdowskie~4, 00-478~Warszawa, Poland, \email{lw@astrouw.edu.pl}
\and
Institute of Astronomy, Faculty of Physics, Astronomy and Informatics, Nicolaus Copernicus University in Toru{\'n}, Grudzi\k{a}dzka 5, 87-100 Toru{\'n}, Poland
\and
Institute of Astronomy, University of Cambridge, Madingley Road, CB3 0HA, Cambridge, UK
\and
Konkoly Observatory, Research Centre for Astronomy and Earth Sciences, H-1121 Budapest, Konkoly Thege út 15–17, Hungary
\and
Astronomical Observatory, Volgina 7, 11060 Belgrade, Serbia
\and
Center for Astrophysics and Cosmology, University of Nova Gorica, Vipavska 11c, 5270 Ajdovščina, Slovenia
\and
Astronomical Institute, University of Wroc{\l}aw, Kopernika 11, 51-622 Wroc{\l}aw, Poland
\and
National Astronomical Research Institute of Thailand (Public Organization), 260 Moo 4, Donkaew, Mae Rim, Chiang Mai 50180, Thailand
\and
 Las Cumbres Observatory, 6740 Cortona Drive, Suite 102, Goleta, CA 93117, USA
\and
School of Physics, Trinity College Dublin, College Green, Dublin 2, Ireland
\and
Vatican Observatory Research Group, Steward Observatory, Tucson, AZ 85721, USA
\and
Flarestar Obsevatory, Fl.5 Ent.B, Silver Jubilee Apt, George Tayar Street, San Gwann, SGN 3160, Malta
\and
Institut de Ci\'encies del Cosmos (ICCUB), Universitat de Barcelona (UB), Mart\'{\i} i Franqu\'es 1, E-08028 Barcelona, Spain
\and
INAF-Osservatorio di Astrofisica e Scienza dello Spazio, Via Gobetti $93/3$, I-40129 Bologna, Italy
\and
Public observatory Astrolab IRIS, Verbrandemolenstraat 5 , 8901 Zillebeke, Belgium
\and
Istanbul University, Faculty of Science, Department of Astronomy and Spaces Sciences, 34119, \.Istanbul T\"urkiye
\and
Istanbul University Observatory Research and Application Center, Istanbul University 34119, \.Istanbul T\"urkiye
\and
Komaba Institute for Science, The University of Tokyo, 3-8-1 Komaba, Meguro, Tokyo 153-8902, Japan
\and
Institute of Earth Systems, University of Malta
\and
ICAMER Observatory, National Academy of Sciences of Ukraine, 27 Acad. Zabolotnoho Str., 03143 Kyiv, Ukraine
\and 
Department of Physics and Astronomy, University of Sheffield, Sheffield, S3 7RH, UK
\and
Landessternwarte, Heidelberg University
\and
Evgeni Kharadze Georgian National Astrophysical Observatory, Abastumani, Georgia
\and
Janusz Gil Institute of Astronomy, University of Zielona Góra, Szafrana 2, 65–516 Zielona Góra, Poland
\and
Faculty of Automatic Control, Electronics and Computer Science, Silesian University of Technology, Akademicka 16, 44-100 Gliwice, Poland
\and
HAO68\_G2-1600
\and
Astronomical Observatory, Jagiellonian University, Orla 171, 30-244 Krak{\'o}w, Poland
\and 
Centre for Advanced Instrumentation, Durham University, UK
\and
Astronomy and Space Physics Department, Taras Shevchenko National University of Kyiv, 4, Glushkova ave., Kyiv, 03022, Ukraine
\and
National Center «Junior academy of sciences of Ukraine», 38-44, Dehtiarivska St., Kyiv, 04119, Ukraine
\and
Dipartimento di Fisica ''E.R. Caianiello'', Universit\'a di Salerno, Via Giovanni Paolo II 132, Fisciano 84084, Italy
\and
INFN, Sezione di Napoli, Via Cintia, Napoli 80126, Italy.
\and 
Millennium Institute of Astrophysics MAS, Nuncio Monsenor Sotero Sanz 100, Of. 104, Providencia, Santiago, Chile
\and Data Science Research Center, Department of Statistics, Faculty of Science, Chiang Mai University, Chiang Mai 50200, Thailand
\and
Zentrum f{\"u}r Astronomie der Universit{\"a}t Heidelberg, Astronomisches Rechen-Institut, M{\"o}nchhofstr. 12-14, 69120 Heidelberg, Germany
\and
Joint Institute for VLBI ERIC, Oude Hoogevceensedijk 4, NL-7991 PD
Dwingeloo, the Netherlands
\and
Institut d'Astrophysique de Paris, Sorbonne Universit\'e, CNRS, UMR 7095, 98 bis bd Arago, F-75014 Paris, France
\and
University of St Andrews, Centre for Exoplanet Science, SUPA School of Physics \& Astronomy, North Haugh, St Andrews, KY16 9SS, United Kingdom
\and
Adiyaman University, Department of Physics, 02040 Adiyaman, Turkey
\and
Astrophysics Application and Research Center, Adiyaman University, Adiyaman 02040, Turkey
\and
Pedagogical University Of Cracow, Podchor\k{a}{\.z}ych 2, 30-084 Krak{\'o}w, Poland
\and
Astronomical Observatory Institute, Faculty of Physics, Adam Mickiewicz University, ul. S{\l}oneczna 36, 60-286 Poznań, Poland
\and
Department of Particle Physics and Astrophysics, Weizmann Institute of Science, Rehovot 76100, Israel
\and
National Astronomical Observatory of Japan, National Institutes of Natural Science, Osawa, Mitaka, Tokyo, Japan 181-8588
\and
Department of Astronomy, University of Virginia, 530 McCormick Rd. Charlottesville, VA 22904, USA
\and
Department of Physics and Astronomy, University of North Carolina at Chapel Hill, Chapel Hill, NC 27599, USA
\and
Departament de Física Qu\'antica i Astrof\'{\i}sica (FQA), Universitat de Barcelona (UB), Mart\'{\i} i Franqu\'es 1, E-08028 Barcelona, Spain
\and
Institut d'Estudis Espacials de Catalunya (IEEC), c. Gran Capit\'a, 2-4, E-08034 Barcelona, Spain
\and
Instituto de Astrof\'isica, Facultad de F\'isica, Pontificia
Universidad Cat\'olica de Chile, Av. Vicu\~na Mackenna 4860, 7820436
Macul, Santiago, Chile\label{inst:Pontificia}
\and
Horten Videregående Skole, Horten, Norway
}

  \date{September 2023}

  \abstract 
{
We present the results of our analysis of Gaia19dke, an extraordinary microlensing event in the Cygnus constellation that was first spotted by the {\gaia} satellite. This event featured a strong microlensing parallax effect, which resulted in multiple peaks in the light curve. We conducted extensive photometric, spectroscopic, and high-resolution imaging follow-up observations to determine the mass and the nature of the invisible lensing object. Using the Milky Way priors on density and velocity of lenses, we found that the dark lens is likely to be located at a distance of $D_L =(3.05^{+4.10}_{-2.42})$~kpc, and has a mass of $M_L =(0.51^{+3.07}_{-0.40}) M_\odot$.
Based on its low luminosity and mass, we propose that the lens in Gaia19dke event is an isolated white dwarf.
}
  
  \keywords{
  Gravitational lensing: micro -- Stars: black holes -- white dwarfs -- Stars: neutron -- Techniques: photometric -- Techniques: spectroscopic }

 \maketitle

\section{Introduction}
In the context of a standard point-source single-lens photometric microlensing event \citep{Paczynski1996}, it is generally challenging to determine a comprehensive set of physical parameters that fully describe the lensing object and its properties. The reason behind this limitation lies in the fact that the standard model of the light curve for such events relies on a single parameter, known as the event's time-scale ($\te$), which is dependent on three physical quantities: the distances of the source and lens, as well as the relative velocity between the lens and the source.

Consequently, it becomes difficult to straightforwardly differentiate between microlensing events caused by main sequence (MS) stars and those caused by stellar remnants like white dwarfs (WD), neutron stars (NS), or stellar-mass black holes (BH) within the vast pool of tens of thousands of photometric microlensing events discovered over the last three decades through dedicated microlensing surveys such as OGLE \citep{Udalski2015}, MOA \citep{MOA2013ApJ...778..150S}, or KMTNet \citep{KMTnet2016JKAS...49...37K}.


The usage of microlensing can be instrumental in shedding light on various unresolved questions concerning stellar remnants, such as the population study and mass distribution of white dwarfs \citep{Raddi2022A&A...658A..22R}, the masses of neutron stars \citep{NS2016ARA&A..54..401O}, the existence of a mass-gap between black holes and neutron stars \citep{Bailyn1998, Ozel2010, Farr:2010}, and the potential of black holes to explain at least a portion of the enigmatic Dark Matter \citep[e.g.][]{Paczynski1986, Wyrzykowski2009, Wyrzykowski2011b, Bird2016, Clesse2015, Carr2018}. 

The chance of unravelling the nature of the lens, hence identification of potential dark lenses, improves in case of microlensing events lasting many months, in contrast to typical events' duration of about one month. 
These long-lasting events often exhibit a phenomenon known as the microlensing parallax effect \cite{Smith2002, Gould2004, GouldMass2004}, which arises due to the Earth's orbital motion around the Sun. This motion leads to a change in the line-of-sight direction, thereby altering the angular separation between the lens and the source. Consequently, the observed amplification undergoes fluctuations that can be characterized by an additional model parameter vector  $\piEvec$, with its components $\piEE$ (East) and $\piEN$ (North).

The length of the vector $\piE$ corresponds to the relative parallax ($\pirel$) between the source and the lens, scaled by the angular size of the event's Einstein Radius ($\thetaE$). It can be expressed as $\piE = (\piL - \piS) / \thetaE$, where $\piL = 1/\DL$ and $\piS = 1/\DS$ correspond to the parallaxes (distances) of the lens and the source, respectively.
Subsequently, the mass and distance of the lensing object can be deduced using the following expressions \citep{Gould2000b, GouldYee2014}:

The mass and distance of the lensing object can then be derived as 
\begin{equation}
M=\frac{\theta_\mathrm{E}}{\kappa \pi_\mathrm{E}} = \frac{\mu_\mathrm{rel} \tE}{\kappa \pi_\mathrm{E}}
\,, ~~~
\kappa \equiv \frac{4G}{c^2 \mathrm{AU}} \simeq 8.1 \frac{\mathrm{mas}}{\msun}
\,,
\end{equation}
and
\begin{equation}
D_{\rm{L}}=\frac{1}{\mu_{\rm{rel}} \tE \pi_{\rm{E}} + 1/D_{\rm{S}}}\, ,
\end{equation}where we used the fact that the angular size of the Einstein radius can be rewritten as a product of the length of the vector of the heliocentric relative proper motion $|\mu_\mathrm{rel}|=|\mu_L-\mu_S|$ 
between the lens (L) and source (S) and the event's timescale $\te$. 

The parallax and time scale are the two physical parameters that can be determined when using the photometric light curves of microlensing events.
Without the knowledge of the Einstein radius ($\thetaE$), the mass and distance of the lens can be determined by employing probability distributions for the density and velocity of lenses \citep[e.g.][]{Wyrz16, WyrzykowskiMandel2020, MrozWyrzykowski2021}. 

One of the methods to obtain $\thetaE$ is to observe both the changes in observed light (photometric component) and the position of the source during the microlensing event (astrometric component).
While a microlensing event occurs, the source is split into two, unevenly magnified images.
Unlike in strong lensing, the angular separation of these images is small and was observed only through the use of Very Large Telescope's instruments GRAVITY and PIONIER \citep{2019DongVLTI, CassanVLTI}.

By obtaining precise measurements of the source's position, it becomes possible to monitor the motion of the light's centroid. This technique is referred to as astrometric microlensing and has demonstrated success in recent observations and discovery of the first isolated stellar-mass black hole with \textit{Hubble} Space Telescope \citep{2022Sahu, 2022Lam, 2022Mroz}.

It will become possible to derive the size of the Einstein radii for many of the brighter microlensing events observed by the European Space Agency's \gaia space mission\citep{Gaia} as \gaia was designed to collect both photometric and astrometric measurements for about 2 billion stars\citep{Gaia, GaiaDR3}. It is anticipated that \gaia's astrometric observations will enable the measurement of astrometric microlensing signals \citep[e.g.][]{Dominik2000, BelokurovEvans2002, Rybicki2018}, which in turn will yield $\thetaE$ \citep{Wyrzykowski2023}. 

To ensure the usefulness of \gaia's astrometry in microlensing events, it was crucial to gather dense and accurate photometric data through intensive monitoring of bright events ($G\lesssim16$) that occurred during the \gaia mission (2014-2025). These events were reported in near-real-time by the \gaia Science Alerts system \citep{Wyrzykowski2012, Hodgkin2013RSPTA.37120239H, Hodgkin2021}.
Of particular significance are the events that exhibit a well-constrained microlensing parallax. When combined with the source distance, these parameters allow for a comprehensive understanding of the lens's distance and luminosity. Consequently, the nature of the lens can be revealed, providing a complete picture of its properties.

In this paper, we present a detailed investigation of the Gaia19dke microlensing event, which satisfies all the aforementioned criteria. The event has already lasted for more than 2000 days, making it a long-duration event. Moreover, it demonstrated a highly pronounced microlensing parallax effect and was sufficiently bright to enable precise astrometry measurements by the \gaia mission. While the \gaia astrometric data will be published in Gaia Data Release 4 ($\sim$Q4 2025), here we present the comprehensive analysis of the photometric data and use the Galactic model to predict the most likely properties of the dark lens. 

The paper is organized as follows. In Section 2 we present the discovery and follow-up observations of the event. Section 3 contains the description of the microlensing model used to fit the photometric data. In Section 4, we analyze the source star using photometry and spectroscopy and in Section 5, we derive the probable parameters of the lens. We discuss the results in Section 6 and conclude in Section 7.

\section{Discovery and follow-up of Gaia19dke}
\begin{figure} 
\centering
\includegraphics[width=8.5cm]{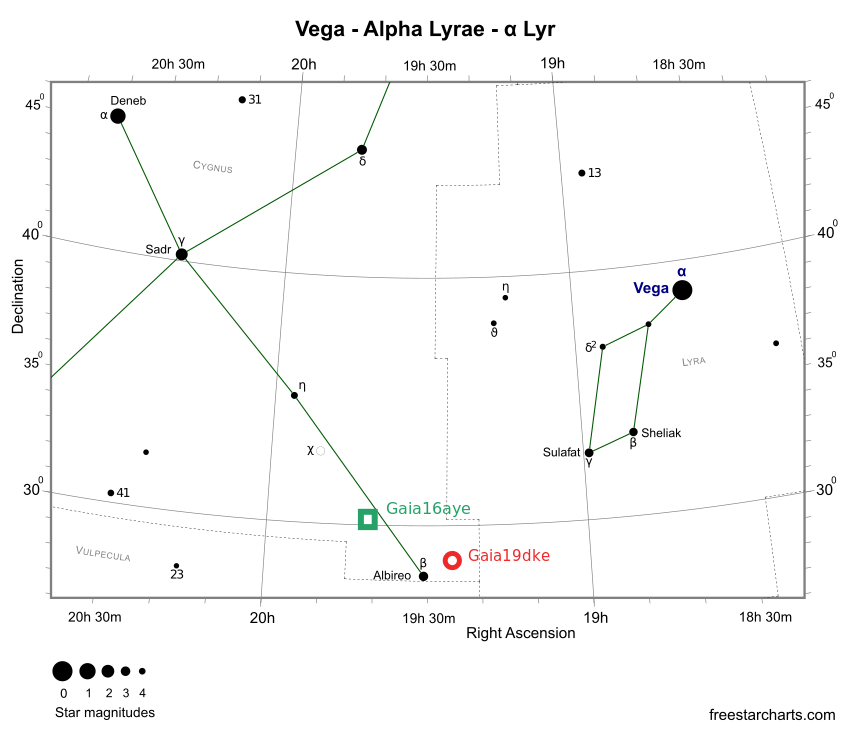}
\caption{Location of the Gaia19dke event (red circle) is shown on the Cygnus - Lyra constellation map from \url{www.freestarcharts.com}. Also shown is the location of Gaia16aye binary microlensing (green square) event from \cite{WyrzykowskiGaia16aye}.
} 
\label{fig:skychart}
\end{figure}

Gaia19dke (IAU Transient Name Server, TNS, id AT2019ndl) event is located in the Cygnus constellation close to the edge of the Lyra constellation (Fig.~\ref{fig:skychart}) in the Northern Galactic Plane ($RA$, $\delta$)=(19:25:58.68, +28:24:24.70) in the equatorial system, ($l$, $b$) = ($62^\circ\hskip-2pt .01113$, $5.\circ\hskip-2pt 70414$ in the Galactic system) 
It was reported by the \gaia Science Alerts System on the 8th of August 2019 (JD' = JD - 2450000. = 8703) as a small rise of brightness in the \gaia G-band in a previously non-varying star. 
\gaia Data Release~3 (\gaia~DR3, \citep{GaiaEDR3Cat}) \texttt{source\_id} is 2026409795566972544. 
The object was previously recorded in the 2MASS catalogue under id 19255869+2824249 \citep{Skrutskie:2006cat}. 

\gaia DR3 for this object provides the following astrometric parameters: 

$\varpi=(0.0718\pm0.0267)$~mas, $\mu_{RA}=(-2.862\pm0.022)$~mas/yr and $\mu_{\delta}=(-5.447\pm0.029)$~mas/yr, where $\varpi$ is the stellar parallax of the source, and $\mu_{RA}$ and $\mu_{\delta}$ are proper motion components in right ascension and declination directions respectively measured at the reference epoch year 2016.

\subsection{{\Gaia} photometry}

While \gaia scans the sky, it revisits the same location on average within 30 days. 
Each transit typically provides two independent measurements separated by 106 minutes coming from the two fields of view of the spacecraft (see \citealt{Gaia} for details). 
As of May 2023, \gaia has collected 191 measurements of Gaia19dke.
The light curve from \gaia is collected in the \gaia broad-band filter $G$-band and exhibits multiple peaks, with the main peak reaching about 14.8~mag in August 2020.

A table with photometric data gathered by \gaia can be found in Table \ref{tab:photGaia}. 
GSA does not provide uncertainty on magnitudes in light curves for published events. 
We, therefore, used {\Gaia} DR3 photometric time-series statistics (mean $G$-band magnitude and its standard deviation) to derive the mean expected uncertainties as a function of magnitude.
The nominal error for the Gaia19dke magnitude range was computed as around 0.008~mag \citep{GaiaDR2}. 
Table \ref{tab:photGaia} presents the uncertainty estimated for {\Gaia} measurements, which were used throughout this work. 

\subsection{Ground-based photometric follow-up}
Due to the fact that the event was relatively bright with $G\sim$15.5 mag at the baseline, it was possible to collect a vast number of follow-up observations using small-sized telescopes. 
The ground-based observations were carried out by a network of telescopes, including manually and robotically operated ones, listed in Table \ref{tab:observatories}.
To facilitate the coordination of observations and data processing, a web-based system called the Black Hole Target and Observation Manager (BHTOM\footnote{\url{https://bhtom.space}}) was utilized, which is based on LCO's Target and Observation Manager TOM) Toolkit \citep{VolgenauTOM2022}.

For each telescope, the acquired images underwent bias, dark, and flat calibration following 
each telescope's procedures and the calibrated fits images were uploaded in near-real-time to BHTOM. PSF photometry was performed using CCDPhot (e.g. \citealt{2020CPCS2, RybickiGaia19bld}, while standardization was achieved using the Cambridge Photometric Calibration Server (CPCS), as detailed in \citep{2019CPCS2}.
Observations were conducted across various filters in both the SDSS and Johnson-Kron-Cousins systems. To establish uniformity, the data were standardized to the Gaia Synthetic Photometry (GaiaSP) catalogue \citep{GaiaSP2023}, with automated matching of instrumental data to the closest filter available in GaiaSP.

Table \ref{tab:photometrystats} lists the number of data points collected by each observatory, the time span of their data and the list of GaiaSP filters the observations were matched to. 
The table contains also the details on the data collected serendipitously for this target by the Zwicky Transient Factory (ZTF) Survey \citep{2019ZTF} in $g$ and $r$ bands and provided by the IPAC service.

The earliest follow-up started 21 days after the announcement of the event on the GSA web page. 
The first data point was taken on the night of 29/30 August 2019, with the 60 cm telescope in the Astronomical Station Vidojevica
(ASV) of Astronomical Observatory, Serbia. 
The follow-up then continued for over 2000 days until the event reached the baseline level again around May 2023.
The data obtained by the follow-up network are available for download from BHTOM page for Gaia19dke (https://bhtom.space). In total, nearly 3000 data points were collected with the telescope network over a period of nearly 4 years.

\subsection{Spectroscopic follow-up}
In order to classify the object and to derive the properties of the source, Gaia19dke was also observed spectroscopically. The first spectrum was obtained close to the first brightness peak on December 11, 2019, with the Spectrograph for the Rapid Acquisition of Transients (SPRAT, \citealt{SPRAT2014})
mounted on 2-m robotic Liverpool Telescope 
(LT, \citealt{2004SteeleLT}) 
located in La Palma, Canary Islands, Spain. The spectrum was taken in the optical part of the electromagnetic window (400-800~nm) and low-resolution mode (R$\sim$350). 
It was reduced, and wavelength and flux were calibrated in a standard way by using an automated pipeline provided by the LT Team. The Xenon arc lamp was used to calibrate the spectrum in the wavelengths. 

SPRAT data have shown the typical spectrum for normal G-type stars with prominent Mg 5167-5184 $\AA$ lines and Balmer series in absorption. No clear emission lines were registered, therefore, we do not observe any hints of stellar activity, variability, or the existence of circumstellar matter. Any of the features responsible for that was not registered in the SPRAT spectrum. Therefore, Gaia19dke was classified as a microlensing event candidate and further follow-up observations were planned.

The Microlensing Observing Platform \footnote{\url{https://mop.lco.global}} automatically requested the spectroscopic monitoring for this target and a low-resolution spectrum (R$\sim$500) has been collected by the OMEGA collaboration on August 8, 2020 (the source was magnified by a factor 1.8 at this time, i.e. $G=14.9$ mag), with the FLOYDS instrument mounted on the Las Cumbres Observatory 2-m telescope at the Siding Spring  observatory \citep{Brown2013}. The spectrum has been reduced with the LCO FLOYDS pipeline\footnote{\url{https://lco.global/documentation/data/floyds-pipeline/}}. It confirmed the classification made based on SPRAT data showing absorption lines typical for a G-type star.

\begin{figure}
\centering
\includegraphics[width=1.1\hsize]{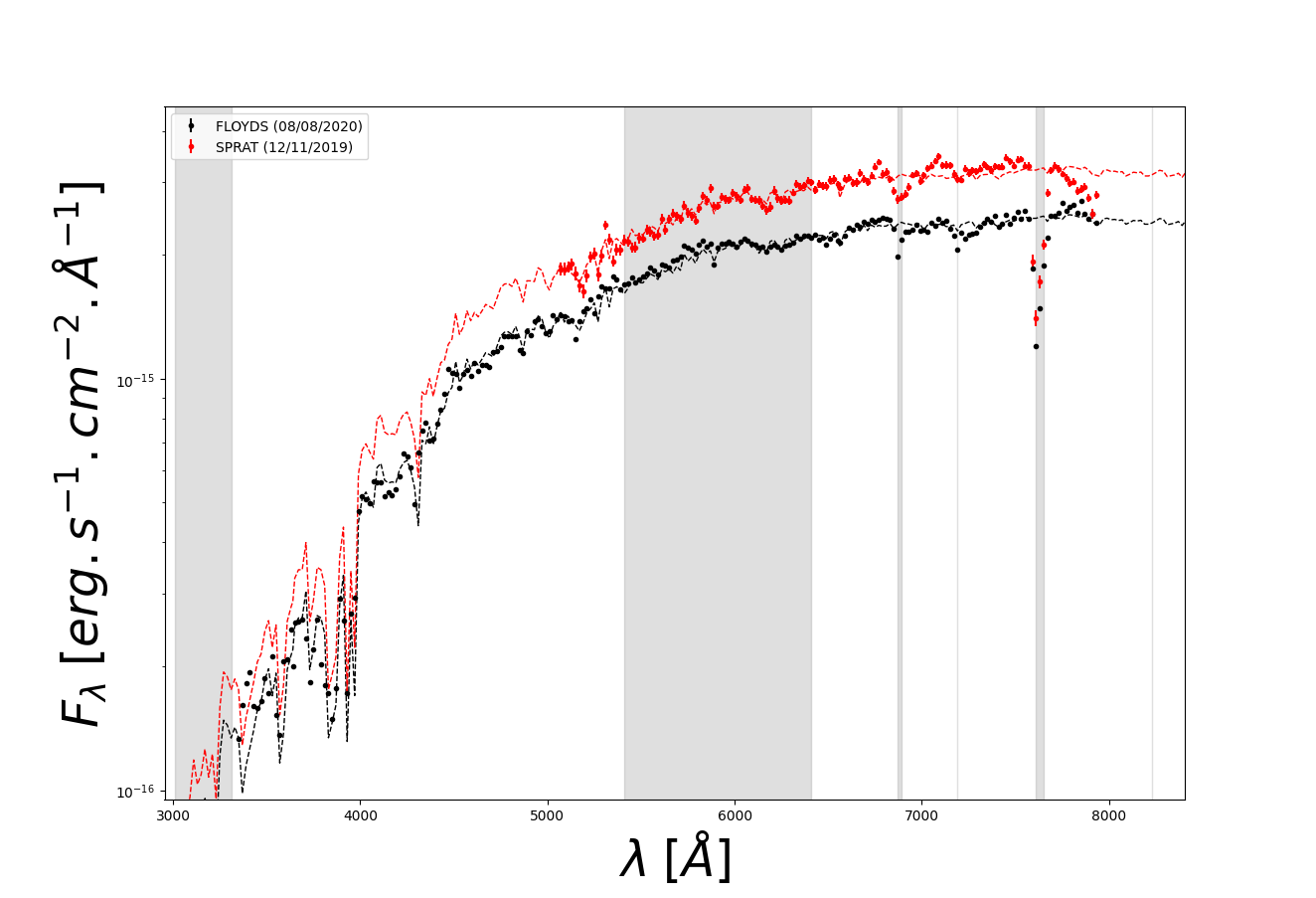}
\caption{Low-resolution spectra of Gaia19dke obtained by LT/SPRAT (red points) and LCO/FLOYDS (black points) spectrographs. The grey parts of the plot denote the wavelength range with the telluric lines. The dashed lines correspond to the best-matching template spectra.
} 
\label{fig:speclowres}
\end{figure}

The low-resolution spectra of Gaia19dke gathered by SPRAT and FLOYDS instruments are presented together in Fig.~\ref{fig:speclowres}.

Gaia19dke event reached a bright enough magnitude near its main peak around mid-2020 to be also observed with high-resolution spectroscopy. We used the Potsdam Echelle Polarimetric and Spectroscopic Instrument \citep[PEPSI, ][]{strass2015} installed at the 2x8.4-m Large Binocular Telescope (LBT)\footnote{\url{https://www.lbto.org/}} located on Mt. Graham, Arizona, US. The data were taken on July 18, 2020, \ie~close to the maximum brightness of the event. The fibre diameter $300~\mu$m as well as two cross-dispersers (CD) were used: III (blue arm) and V (red arm) simultaneously. We were able to obtain a high-dispersion spectrum with an S/N ratio of around 31 and resolution R$\sim$43\,000, which covers the wavelength range $383-907$~nm. It was calibrated by using the standard PEPSI software for stellar spectroscopy (SDS4PEPSI, \citep{2000Ilyin}), \ie~images were bias subtracted, flat-fielded, and then optimally extracted and normalized using a spline fit to the continuum.
Due to the poor quality of the spectrum below 480~nm, for further analysis, we used part above this threshold.

\begin{figure}
\centering
\includegraphics[width=0.5\textwidth]{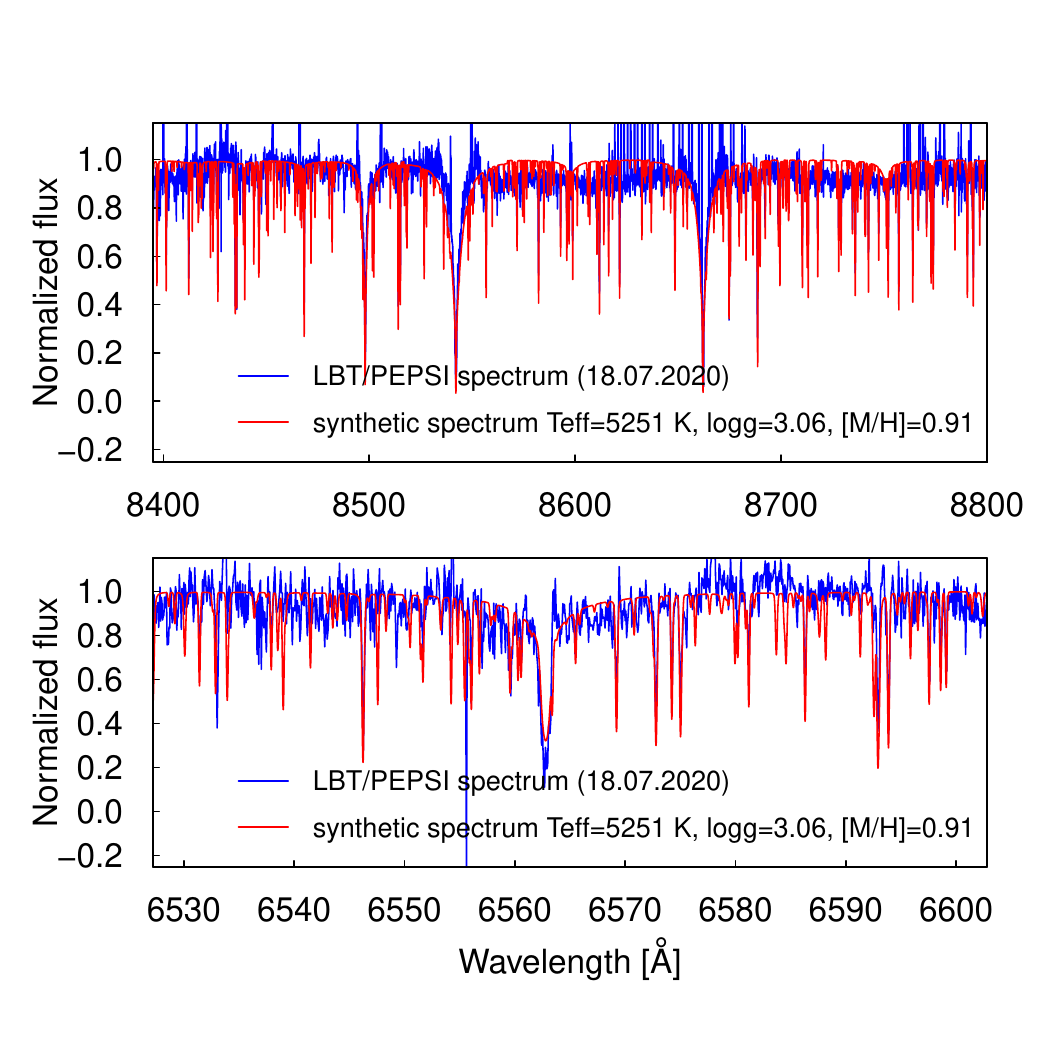}
\caption{Spectrum of the Gaia19dke obtained on 18 July 2020 with LBT/PEPSI around the main peak of the event({\it blue}) and the best-matching fit ({\it red}) synthesized for the specific parameters. The Ca~II triplet ({\it top}) and H$\alpha$ ({\it bottom}) region are presented.}
\label{fig:speclbt}
\end{figure}

The spectrum from a high-resolution PEPSI spectrograph is presented in Fig.~\ref{fig:speclbt}. In addition, the synthetic spectrum ({\it red}) generated based on the method described in Section~4 is over-plotted on the observed spectrum ({\it blue}).

\subsection{High-resolution imaging follow-up}
\label{sec:highres}

Gaia19dke was observed with the Gemini North 8-m telescope using the ‘Alopeke speckle imaging instrument\footnote{\url{https://www.gemini.edu/sciops/instruments/alopeke-zorro/}} on 9 August 2020. ‘Alopeke is a simultaneous two-channel EMCCD instrument that performs speckle interferometric imaging. Using narrow-band filters centred at 562 nm and 832 nm, the images are obtained with 60 msec integration times and collected in sets of 1000 such images/set.  The final product from ‘Alopeke imaging is a high-resolution image in each filter with an inner working angle at the diffraction limit, near 20 mas for the 8-m Gemini telescope, and covering a small field of view out to 1.2~arcsecs. 

\begin{figure}
    \centering
    \includegraphics[width=0.5\textwidth]{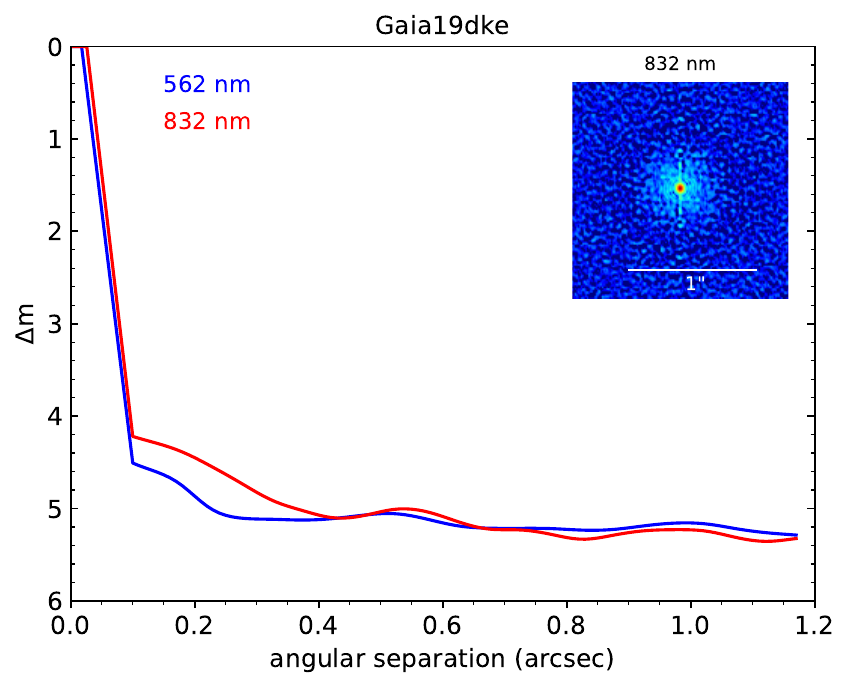}
    \caption{Contrast curves for red and blue narrow-band filters obtained from speckle interferometric observations of Gaia19dke obtained on 2020 Aug.9 with 'Alopeke instrument at the Gemini telescope. The inset shows the combined set of images in an 832~nm filter.
    }
    \label{fig:astrometry}
\end{figure}

The set of images was subjected to Fourier analysis in our standard reduction pipeline \citep{Howell2011}. Figure \ref{fig:astrometry} shows the final 5-$\sigma$ contrast curves in each filter and the 832~nm reconstructed speckle image. We find that the object at Gaia19dke is not resolved beyond a single point source, even down to the 20 mas inner working angle. 

\section{Photometric Microlensing Model} \label{sec:ulens_model}

The photometric data of Gaia19dke has been modelled with the single point source single lens microlensing model with annual parallax\citep[e.g.][]{Gould2000b, Smith2002, Wyrz16, RybickiGaia19bld, Kruszynska2022}.
We used open-source flexible software \textit{MulensModel}\citep{PoleskiMulensModel} for finding the model parameters.

The parallax model is described with the following parameters:
\begin{itemize}
    \item $t_0$, time of the minimal approach between the lens and the source;
    \item $u_0$, impact parameter, the minimal distance between the lens and the source in units of the Einstein Radius;
    \item $t_E$, the time-scale of the event, defined as the time to cross the Einstein Radius;
    \item $\vec{\pi_E}$, vector of the microlensing parallax, decomposed into equatorial North $\pi_{EN}$ and East $\pi_{EE}$ components;
    \item $mag_0$, baseline magnitude(s), separately in each observing band, computed in \textit{MulensModel} from source flux;
    \item $f_S$, blending parameter(s), separately in each observing band, defines as the flux of the source over the total baseline flux, composed of source and blend(s) and/or lens light, computed in \textit{MulensModel} from source and blend fluxes;
\end{itemize}

The microlensing parallax model has been fitted in a geocentric frame with a fixed $t_{0_{par}}$ parameter, set to the time of the maximum of the light curve, hence very close to $t_0$. 
To find the most likely model, we used Markov chain Monte Carlo (MCMC) implemented in the \texttt{emcee} package \citep{EMCEE}. 
Since {\gaia} observed Gaia19dke from L2 point, we included the space-parallax factor in \textit{MulensModel}.

\begin{figure*} 
\centering
       \includegraphics[width=\hsize]{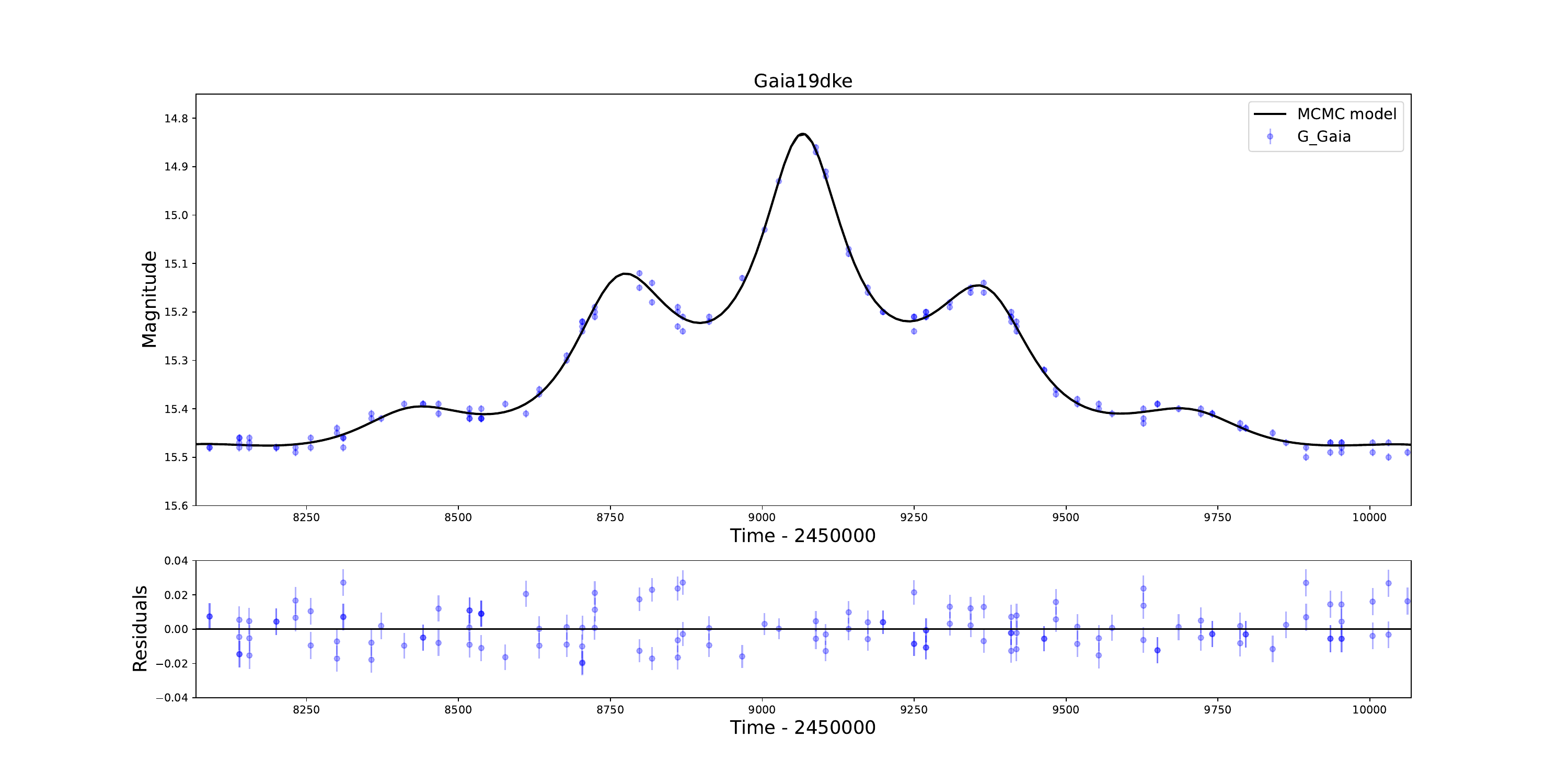}
\caption{Light curve of Gaia19dke microlensing event with data only from {\gaia}, spanning from JD = 2458062 to JD = 2460062. The black line is the mode of the chains from the MCMC model. The bottom panel shows the residuals with respect to the mode solution.
} 
\label{fig:lcgaia}
\end{figure*}

We used all photometric light curve data gathered by the end of May 2023, when the event reached its baseline magnitude.
We first modelled {\gaia} data only, as it covers the shape of the event densely and contains a couple of years of the baseline prior to the microlensing event.
Table \ref{tab:model} contains the values of microlensing model parameters found when fitting {\gaia}-only data. Our procedure identified only one solution in the parameter space for negative $u_0$. 
Figure \ref{fig:lcgaia} shows the {\gaia} photometric data together with the best microlensing model with a parallax fit to that data.

Subsequently, the microlensing model fitting was performed using the combined dataset of {\gaia} observations, follow-up observations, and data from the Zwicky Transient Facility (ZTF). Given that all observations, acquired with a network of telescopes, underwent consistent calibration and standardization to GaiaSP bands, we were able to effectively utilize the entire collected dataset from all telescopes, a total of nearly 5000 data points. 
However, we excluded 30 points calibrated to $u$, $U$, and $z$ filters, as they were erroneously matched to incorrect bands and exhibited clear outliers.
The modelling has been carried out in each GaiaSP band separately. 

\begin{figure*} 
\centering
\includegraphics[width=\hsize]{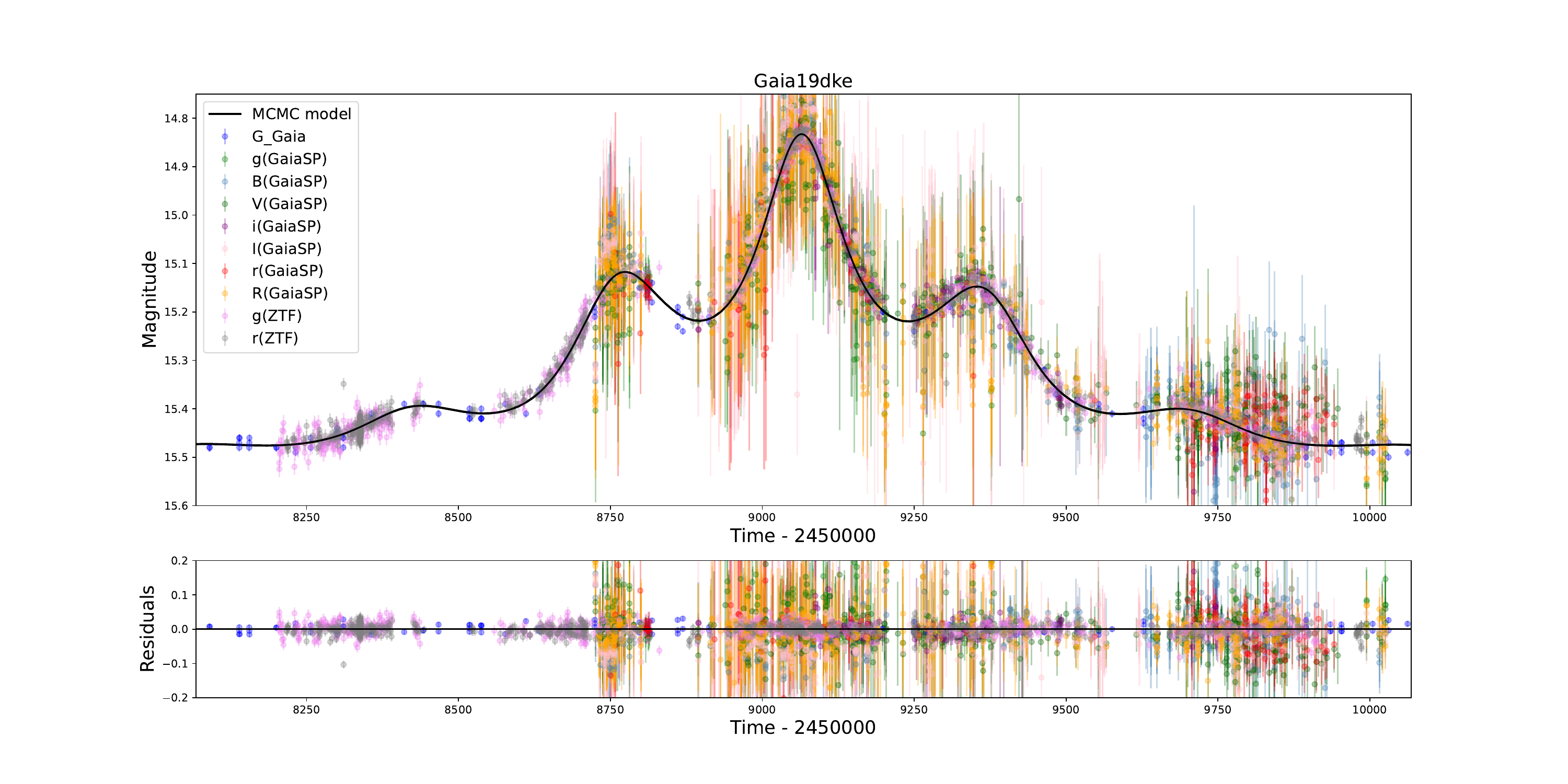}
\caption{Light curve of Gaia19dke microlensing event with data from {\gaia} and follow-up observations, spanning from JD = 2458062 to JD = 2460062. Black line is the mode of the chains from the MCMC model. The bottom panel shows the residuals with respect to the mode solution.
} 
\label{fig:lcfup}
\end{figure*}

Table \ref{tab:model} shows the values of microlensing model parameters found for {\gaia} and the follow-up data set combined. The baseline magnitude and blending parameters were found separately for each observatory and filter. 
There was also only one parallax solution found for this data set.
Figure \ref{fig:lcfup} shows the best microlensing model and its residuals fitting the {\gaia} and follow-up observations. 

The parameters obtained in the two models agree within the margin of error, but the model constructed using follow-up photometric observations exhibits narrower error bars, a factor of 3 to 5 better, which translates to more precise parameter estimates and improved accuracy. In order to achieve more continuous samples from the parameter space, in the modelling process we allowed the blending parameter $f_s$ to be greater than one. Samples with $f_s$ greater than one should be treated as if there is no blending at all. In both models, the value of the blending parameter is very close to 1, in particular, for $\Gaia$ band, $f_s=1$ within the margin of error. Other bands yielded slightly lower values of $f_s$ (e.g. I(GaiaSP)), which can be attributed to low spatial resolution of instruments collecting these data and the observed blending is caused by nearby stars falling within their disks of Point Spread Function. 


\begin{figure} 
\centering
\includegraphics[width=\hsize]{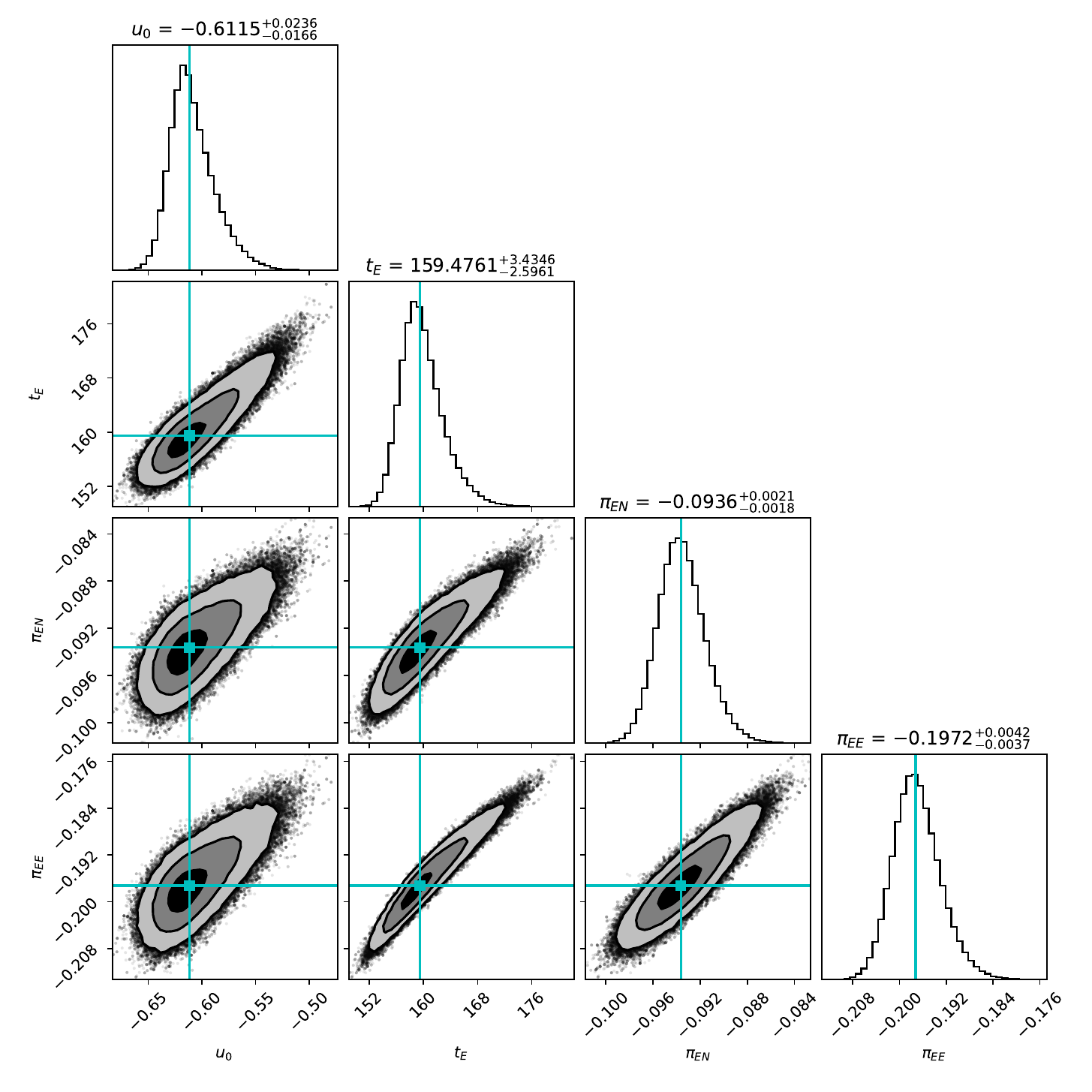}
\caption{
Chi-squared contours plotted as a function of the parameters fitted in the MCMC fit for the best model for the Gaia19dke event obtained with \gaia-only data. Black, dark grey and light grey solid colours represent $1 \sigma$, $2 \sigma$, and $3 \sigma$ confidence regions respectively. Black dots represent solutions outside of the $3 /sigma$ confidence level. Cyan lines and squares mark the median solution reported in Table \ref{tab:model}. 
The plot has been created using \texttt{corner} python package by \cite{corner}.
} 
\label{fig:cornergaia}
\end{figure}

\begin{figure} 
\centering
\includegraphics[width=\hsize]{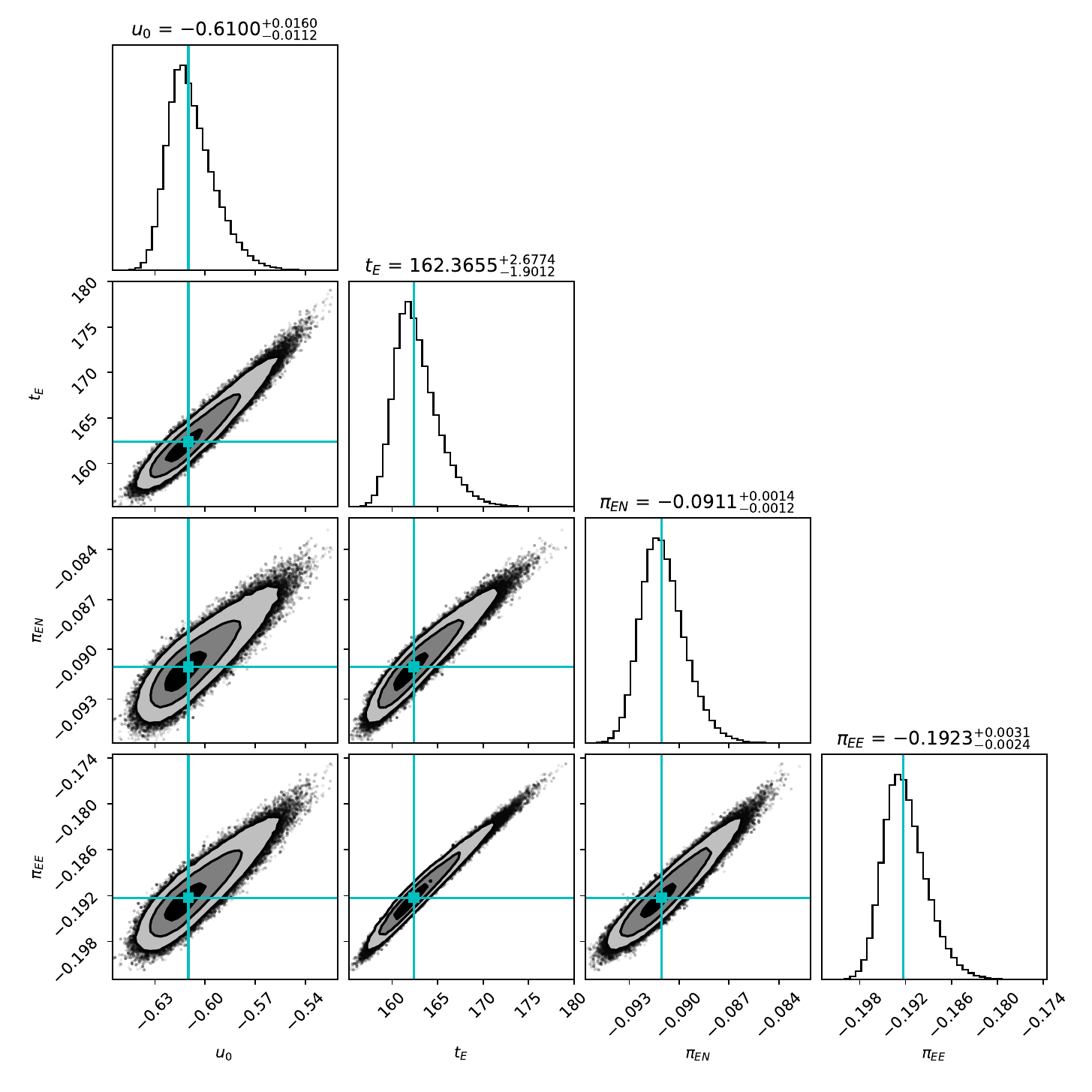}
\caption{
Chi-squared contours plotted as a function of the parameters fitted in the MCMC fit for the best model for the Gaia19dke event obtained after including the follow-up data. Black, dark grey and light grey solid colours represent $1 \sigma$, $2 \sigma$, and $3 \sigma$ confidence regions respectively. Black dots represent solutions outside of the $3 \sigma$ confidence level. Cyan lines and squares mark the median solution reported in Table \ref{tab:model}. 
The plot has been created using \texttt{corner} python package by \cite{corner}.
} 
\label{fig:cornerfup}
\end{figure}

\begin{table} 
\centering
\caption{Microlensing parallax model for {\gaia}-only data and {\gaia} with follow-up observations.
}
\label{tab:model}
\begin{tabular}{lll}
\hline
\noalign{\smallskip}
Parameter  &  {\gaia}-only & {\gaia}+FUP \\
\noalign{\smallskip}
\hline
\hline
\noalign{\smallskip}
$t_{0,par}-2450000.$ [JD] & - & 9068 \\ \noalign{\smallskip}
$t_0-2450000.$ [JD] & 9065.39$^{+0.82}_{-0.81}$ & 9064.0639$^{+0.33}_{-0.33}$   \\ \noalign{\smallskip}
$\tE$ & 159.48$^{+3.43}_{-2.60}$ & 162.47$^{+2.68}_{-1.90}$\\
\noalign{\smallskip}
$u_0$ & -0.6115$^{+0.0236}_{-0.0166}$ & -0.6100$^{+0.0160}_{-0.0112}$ \\
\noalign{\smallskip}
$\piEN$ & -0.0936$^{+0.0021}_{-0.0018}$ & -0.0911$^{+0.0014}_{-0.0012}$\\
\noalign{\smallskip}
$\piEE$ & -0.1972$^{+0.0042}_{-0.0037}$ & -0.1923$^{+0.0031}_{-0.0024}$\\
\noalign{\smallskip}
$mag_0$ $G$ (Gaia) & 15.5052$^{+0.0007}_{-0.0006}$ & 15.5059$^{+0.0005}_{-0.0004}$ \\
\noalign{\smallskip}
$\fs$ $G$ (Gaia) & 1.0045$^{+0.0437}_{-0.0604}$ & 0.9947$^{+0.0301}_{-0.0421}$ \\
\noalign{\smallskip}
$mag_0$ $B$(GaiaSP) & - & 17.2458$^{+0.0007}_{-0.0006}$ \\
\noalign{\smallskip}
$\fs$ $B$(GaiaSP) & - & 0.9162$^{+0.0280}_{-0.0397}$  \\
\noalign{\smallskip}
$mag_0$ $g$(GaiaSP) & - & 16.6105$^{+0.0014}_{-0.0014}$ \\
\noalign{\smallskip}
$\fs$ $g$(GaiaSP) & - & 0.9832$^{+0.0303}_{-0.0426}$  \\
\noalign{\smallskip}
$mag_0$ $i$(GaiaSP) & - & 14.9657$^{+0.0298}_{-0.0419}$ \\
\noalign{\smallskip}
$\fs$ $i$(GaiaSP) & - & 0.9685$^{+0.0298}_{-0.0419}$  \\
\noalign{\smallskip}
$mag_0$ $I$(GaiaSP) & - & 14.4646$^{+0.0008}_{-0.0007}$ \\
\noalign{\smallskip}
$\fs$ $I$(GaiaSP) & - & 0.8624$^{+0.0265}_{-0.0375}$  \\
\noalign{\smallskip}
$mag_0$ $r$(GaiaSP) & - & 15.4483$^{+0.0010}_{-0.0010}$ \\
\noalign{\smallskip}
$\fs$ $r$(GaiaSP) & - & 0.9324$^{+0.0289}_{-0.0409}$  \\
\noalign{\smallskip}
$mag_0$ $R$(GaiaSP) & - & 15.2199$^{+0.0007}_{-0.0007}$ \\
\noalign{\smallskip}
$\fs$ $R$(GaiaSP) & - & 0.9849$^{+0.0301}_{-0.0427}$  \\
\noalign{\smallskip}
$mag_0$ $V$(GaiaSP) & - & 15.9731$^{+0.0007}_{-0.0007}$ \\
\noalign{\smallskip}
$\fs$ $V$(GaiaSP) & - & 0.9987$^{+0.0306}_{-0.0433}$  \\
\noalign{\smallskip}
$mag_0$ $g$(ZTF) & - & 16.5351$^{+0.0009}_{-0.0009}$ \\
\noalign{\smallskip}
$\fs$ $g$(ZTF) & - & 0.9690$^{+0.0297}_{-0.0421}$  \\
\noalign{\smallskip}
$mag_0$ $r$(ZTF) & - & 15.3947$^{+0.0009}_{-0.0008}$ \\
\noalign{\smallskip}
$\fs$ $r$(ZTF) & - & 0.9777$^{+0.0290}_{-0.0404}$  \\
\noalign{\smallskip}
$\chi^2$ & 556.7 & 3621.64  \\
\noalign{\smallskip}
\hline
\hline
\end{tabular}
\end{table}

\section{Source star} \label{sec:source_star}

In order to determine the parameters of the lensing object, the initial step involves deducing the distance and the spectral type of the source star. 

Our study is based on the assumption that the source star is single since there are no signs of its binarity in the microlensing model. Moreover, according to {\gaia} EDR3, the closest object is 1.6 arcsecs away and is significantly fainter.

\subsection{Atmospheric parameters}
The parameters of the source star in the Gaia19dke event were derived from spectroscopic follow-up datasets, high-resolution data obtained with LBT/PEPSI and low-resolution data from two instruments: LT/SPRAT and LCO/FLOYDS.

The spectroscopic analysis of absorption lines visible in high-resolution PEPSI spectrum was performed first. We used {\it iSpec}\footnote{\url{https://www.blancocuaresma.com/s/iSpec}} framework for spectral analysis which integrates several well-known radiative transfer codes \citep{BlancoCuaresma2014, BlancoCuaresma2019}. In our case, to determine atmospheric parameters (i.e., effective temperature $T_{\rm eff}$, surface gravity $\log g$, metallicity [M/H], microturbulence velocity $v_{\rm t}$), the SPECTRUM\footnote{\url{http://www.appstate.edu/~grayro/spectrum/spectrum.html}} code was used. We generated a set of synthetic spectra based on a well-known grid of MARCS atmospheric models \citep{Gustafsson2008} and solar abundances taken from \citet{Grevesse2007}. The synthetic spectra were fitted to the observational spectrum for selected regions containing H$\alpha$, Ca, Mg, Fe, Na, and Ti atomic lines. The best-matching solution was found for the following parameters: $T_{\rm eff} = (5251\pm 25)$~K, $\log g = (3.06 \pm 0.02)$, $\mathrm{[M/H]} = (0.91 \pm 0.03)$~dex and $v_{\rm t} = (1.23 \pm 0.07)$~km~s$^{-1}$. According to these parameters, we assume that our source star is a metal-rich G5-type giant. Moreover, no absorption lines from a potential second component are visible in PEPSI data. Fig.~\ref{fig:speclbt} shows the result of this analysis, \ie~PEPSI spectrum and synthetic fit for Ca~II triplet and H$\alpha$ region are presented.

After that, we modelled the spectroscopic data with templates on the full wavelength range. This approach is complementary to the analysis of absorption lines presented above. Following the method of \citet{Bachelet2022}, we fitted the FLOYDS and SPRAT spectra with templates from \citep{Kurucz1993} with the Spyctres pipeline\footnote{\url{https://github.com/ebachelet/Spyctres}}. The new version of Spyctres includes the updated extinction law from \citet{Cardelli1989} to the one of \citet{Wang2019}. In short, the latter combines an adjustment of the \citet{Cardelli1989} law with a fixed total-to-selective extinction ratio $R_V=A_V/E(B-V)=3.1$ and a power-law index $\alpha=2.07$ for the near-IR regions. The data and results are presented in the Fig.~\ref{fig:speclowres}. The template-matching analysis reveals that the source is a red giant, with an effective temperature $T_{eff}=(5000\pm200)$~K, a sun-like metallicity $[M/H]=(0.0\pm0.3)$~dex, a surface gravity $\log g=(2.2\pm0.5)$, an angular radius $\theta_*=(7.9\pm0.4)~\mu as$ and an absorption $A_v=(1.6\pm0.2)$~mag. 

The results obtained from absorption line analysis and template-matching are in good agreement, except the metallicity, and are presented in Tab.~\ref{tab:params}.

\begin{table}
\centering
\caption{Summary of the derived parameters for the source of Gaia19dke event. Averaged solutions of line fitting and template matching are presented.}
\label{tab:params}
\begin{tabular}{lcc}
\hline
\hline
Parameter           & Line fitting    & Template Matching \\
\hline
$T_{\rm eff}$ [K]   & $5251 \pm 25$   & $5000 \pm 200$ \\
${\log g}$          & $3.06 \pm 0.02$ & $2.2 \pm 0.5$ \\
${\rm [M/H]}$ [dex] & $0.91 \pm 0.03$ & $0.0 \pm 0.3$  \\
$v_{\rm t}$ [km/s]  & $1.23 \pm 0.07$ & -- \\
$A_{\rm v}$ [mag]   &    --           & $1.6 \pm 0.2$  \\
${\rm \theta_*}$ $[\mu as]$ &   --      & $7.9 \pm 0.4$ \\
\hline
\hline
\end{tabular}
\end{table}

\subsection{Source distance}
One of the simplest and most popular ways to determine the distance to the star is to use the \cite{2021Bailer-Jones} catalogue, where distances were calculated based on the {\gaia} EDR3 and priors on the Galaxy. Geometric distance, based on the parallax and its uncertainties, gives the distance to Gaia19dke source star of $7.6 < D_s < 11.9 $~kpc. The photo-geometric value, which is based on the parallax, the colour as well as the observed magnitude of the star, gives the distance of $6.7 < D_s < 9.2$~kpc. We note here, that the values based on Gaia parallax measurement in case of microlensing events should be considered with great care, as the parallax measurement can be affected by the light of the lens, if luminous, or any other blends in the line of sight. Moreover, if the parallax measurement obtained from the astrometric time-series collected at the time of the event, the astrometric data can be also affected by the astrometric microlensing effect (e.g. \citealt{Rybicki2018, 2022Sahu, Jablonska2022}). Therefore, in order to verify the distance to the source star, we use the spectroscopic data and apply the well-known spectro-photometric equation: 

\begin{equation}
5\log D_S = V - M_V + 5 - A_V,
\end{equation}
where $D_S$ is the distance to source star, $V$ is the apparent magnitude, $M_V$ is the absolute
magnitude and $A_V$ is the interstellar extinction. 

In the present work, we used the atmospheric parameters based on the high-resolution PEPSI spectrum where the star is classified as G5 giant, while the extinction value $A_V=1.6\pm0.2$~mag was taken from the template matching analysis of low-resolution spectra. 

The typical absolute magnitude and error for G5 giant star are $M_V= (1.0 \pm 0.5)$~mag \citep{Straizys1992}. Together with the apparent magnitude of Gaia19dke $V=16.101$~mag \citep{StassunTESS} and accepting extinction value $A_V=1.6$~mag determined from the low-resolution spectra and taking into account the non-linearity of the transformation and asymmetry of the distance, $\log_{10} (D_S )$, we have determined the distance to the source star of Gaia19dke $D_s= (4.9 \pm 1.2)$~kpc which is a factor of two different from \citet{2021Bailer-Jones}'s values. It is in good agreement with the template matching analysis that points towards a source distance of $D_s=4.3^{+3.3}_{-1.1}$ kpc assuming a source age of 1 Gyr and using the isochrones from \citet{Bressan2012} and \citet{Marigo2013}. 

Because of the significant difference between our spectroscopic distance and literature values from \cite{2021Bailer-Jones}, we should critically evaluate which value is the most real and which one should be used for determining the lens parameters.




To independently verify the source star parameters we apply other available methods based on accessible databases. We used infrared photometry from 2MASS \citep{Skrutskie:2006cat} survey were source stars measured magnitudes in $J = (13.348 \pm 0.024)$~mag, $H = (12.766 \pm 0.024)$~mag and $K_s = (12.550 \pm 0.023)$~mag.
According to \citep{StraizysLazauskaite}, the intrinsic colour of the G5 giant star should be $(J-K_s)_0 = 0.49$~mag. 
 
 \begin{figure}
\centering
\includegraphics[width=8.5cm]{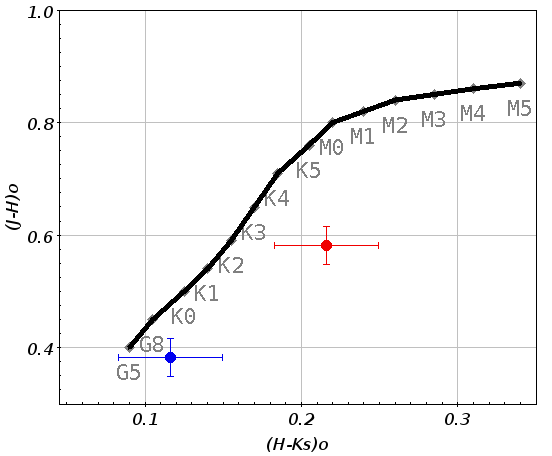}
\caption{Colour-colour $(J-H)_0$ vs. $(H-K)_0$ diagram for the intrinsic red giant's branch (black line). Spectral classes, corresponding to the intrinsic colours, are indicated close to the line. The value for  Gaia19dke is plotted as a red point with errors. De-reddened and shifted according to an extinction value $A_{K_s} = 0.21$~mag is shown as a blue point with errors.} 
\label{fig:2mass}
\end{figure}
 
For the source star, the colour excess and interstellar extinction were calculated with the following equations \citep{Dutra}:
\begin{equation}
E_{J-K_S} = (J-K_s)_{\rm obs} - (J-K_s)_0, \\
A_{K_s} = 0.67\,E_{J-K_S},  
\label{eq:ak}
\end{equation}

where $E_{J-K_S}$ is the colour excess, $(J-K_s)_{\rm obs}$ is the observed colour, $(J-K_s)_0$  is the the intrinsic colour, and $A_{K_s}$ is the interstellar extinction in $K_s$ band.

 According to Eq. \ref{eq:ak}, the estimated extinction value for this star is $A_{K_s} = 0.21$~mag. The extinction value $A_{K_s}$ transformed to the $A_V$ with the following relation \citep{Cardelli, Dutra}: 
 \begin{equation}
A_V = 8.3 A_{K_s},    
\label{eq:av}
\end{equation}
The estimated value of $A_V = 1.7 \pm0.3$~mag is in excellent agreement with the value determined by template-matching based on low-resolution spectra $A_V=1.6\pm0.2$~mag. Fig. \ref{fig:2mass} shows the location of the source star in the 2MASS $(J-H)_0$ vs. $(H-K)_0$ diagram for the observed and dereddened according to an extinction value. The intrinsic red giant's branch is shown as a black line. The dereddened star position on the diagram shows acceptable agreement with extinction and spectral class determined based on low- and high-resolution spectra collected for Gaia19dke. 


We used another method that allows us to verify extinction was proposed by \citep{Majewski2011ApJ} based on combined 2MASS and Spitzer colour indices $H$-[4.5], since for most of F-G-K stars are close to the zero. Here [4.5] is the magnitude at 4.5 $\mu$m of the Spitzer IRAC system. We have to apply the WISE \citep{Wright2010AJ} system since the Spitzer measurements are absent and taking into account that WISE $W2$ measurements with the 4.6 $\mu$m mean wavelength direct comparison  \citep{Jarrett2011ApJ} shows little scattering. Using WISE  measured magnitude $W2 = (12.524 \pm 0.027)$~mag, for the source star, interstellar extinction was calculated with the equation:   

\begin{equation}
A_{K_s} = 0.918\,(H-W2-0.08),    
\end{equation}

In this way, the estimated extinction value $A_{K_s} = 0.149$~mag is by 0.06~mag smaller than previously determined using only 2MASS.  

The answer seems obvious that extinction value $A_V=1.6$~mag determined by the spectroscopic analysis and compared with different methods matches with calculated using different databases.  

For distance check, we also use 2MASS photometry. We again apply the spectro-photometric method but use 2MASS $K_s$ where distance is determined with the following equation:

\begin{equation}
5\log D_S = K_s - M_{K_s} + 5 - A_{K_s},
\label{eq:ds}
\end{equation}

The most uncertain in Eq.\ref{eq:ds} are $M_{K_s}$ for the type G5 giants. We assume its value of $-1.5$~mag since the location in $M_{K_s}$/$J-K$ HR diagram is on the left edge from the Red Clump Giant (RCG) position \citep{Veltz}. We do not exclude that the real $M_{K_s}$ may vary more than $\pm 0.5$~mag. Using 2MASS photometry we just verify the distance and we determine $D_s = (6.0 \pm 1.4)$~kpc to the source star.

As demonstrated above, spectro-photometric method based on optical and infrared data yielded a similar value in the source distance as the one using spectra. We assume that the optically determined distance is more reliable than the infrared one because, in the 2MASS colour-colour diagram, the star only coincides with the actual position of the G5 giant within the error limits, which can be explained by the measurement errors. We can not exclude some variability properties \citep{Henry2000ApJS} since it can change observed magnitude and colour, consequently and source star location on 2MASS colour-colour diagram. 

Throughout the work, we, therefore, use the source distance determined with the PEPSI spectrum, $D_s= (4.9 \pm 1.2)$~kpc.

\section{Lensing object}
\label{sec:lens}



The microlensing model found for Gaia19dke (Section \ref{sec:ulens_model}) indicates no additional light in the event apart from the source. This is encompassed in the blending parameters derived for each photometric band, as listed in Table \ref{tab:model}.
Blending can originate from both the lens itself as well as any star located in close vicinity of the event and unresolved by the photometry. Gaia19dke is located in the Galactic Disk, where the stellar density is significantly lower than in typical microlensing fields in the Galactic Bulge, hence we do not expect any additional source of light close to it, which is confirmed with the high-angular resolution imaging with 'Alopeke (Sec.\ref{sec:highres}).

In order to constrain the nature of this dark lensing object, hence its mass and distance, we adopted the method outlined in \citep{Wyrz16}, \citep{MrozWyrzykowski2021}, \citep{Kruszynska2022}, \citep{VVVdarklenses} and explained in detail in Howil et al. (in prep.). 
The microlensing parameters and their samples from MCMC obtained in previous steps, described in Section \ref{sec:ulens_model}, were combined with priors on the mass, distance, and velocity distribution of stars in the Galaxy for the lens and the source.
Blending parameters $\fs$ of for both {\gaia}-only and {\gaia} with follow-up are close to 1, which means {\gaia} registers the movement and position of the source star.
We have thus adopted the proper motion for the source star as published in {\gaia}~EDR3.
For the distance, we used the value obtained from spectral analysis, described in Section \ref{sec:source_star}. 
In each iteration, we have drawn from a Gaussian distribution of distances with a mean of $4.9\,\text{kpc}$ and a spread of $1.2\,\text{kpc}$.
This method requires also knowing the value of the extinction $A_\mathrm{G}$ towards the lens, to constrain the light coming from the lens if it was an MS star.
We used the value presented in {\gaia}~DR2 catalogue, which lists $A_\mathrm{G}$ under \texttt{a\_g\_val} in \texttt{gaia\_source} table and is equal to $A_\mathrm{G} = 0.8043$~mag. 
We assume this value to be the maximal possible extinction in the direction towards the lens.
Finally, we had to assume the relative proper motion of the lens and source $\murel$. 
For this, we drew a random number between 0 and 30~mas $\mathrm{year}^{-1}$ \citep{MrozWyrzykowski2021}.
This allowed us to find the distance and mass to the lens in combination with the $\piE$ and $\tE$ obtained from the posterior distribution of parameters of the best-fitting microlensing model solution and the distance mentioned above to the source.
Knowing the mass and distance of the lens, we could derive the observable brightness of the lens as if it was the MS star using empirical data from \cite{PecautMamajek2013}\footnote{\href{http://www.pas.rochester.edu/~emamajek}{http://www.pas.rochester.edu/\~{}emamajek}} and compare it to the constraints on the brightness of the lens we obtained from microlensing model.
We then computed a weight using a set of priors from \citet{Skowron2011} for all the pairs of lens mass $\ML$ and lens distance $\DL$.
For the mass function prior we used the value of -2.35, following the classical mass function for stars \citep{Kroupa2003}.

\begin{figure}
    \centering
    \includegraphics[width=10.0cm]{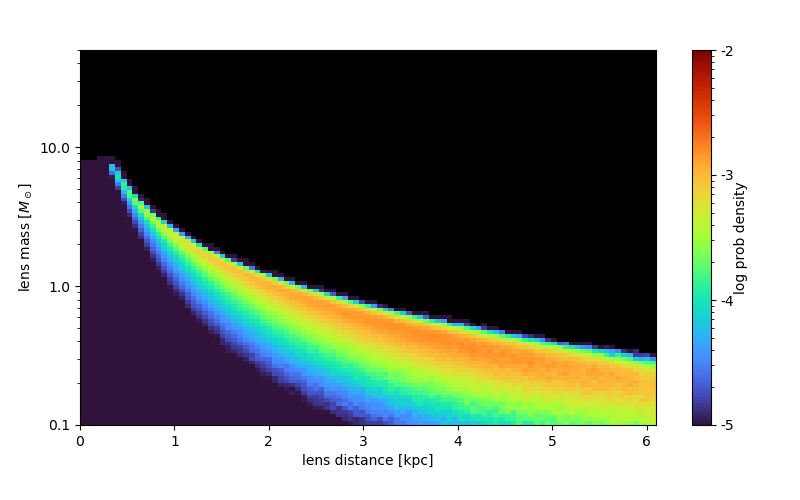}
    \caption{
    Dark Lens code output for Gaia19dke microlensing solution. Lens mass-distance probability density.}
    \label{fig:mass_distance_dlc}
\end{figure}

\begin{figure}
    \centering
    \includegraphics[width=10.0cm]{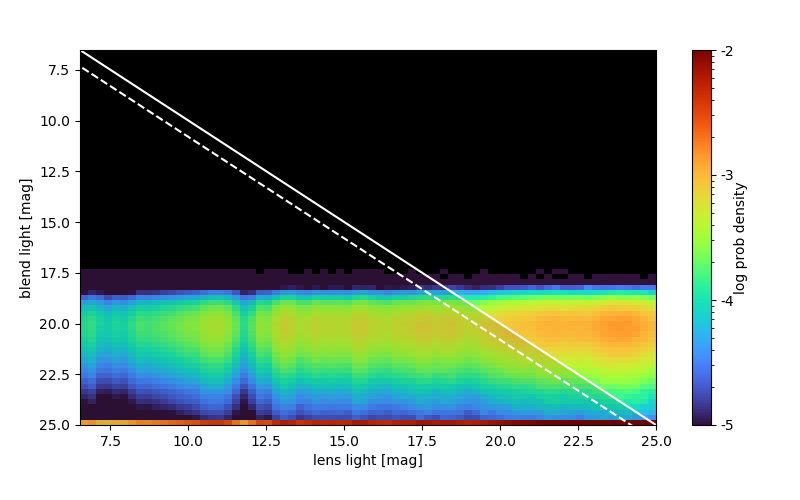}
    \caption{
    Dark Lens code output for Gaia19dke microlensing solution. G-band blend light vs lens light probability density. The lines divide between the case where the lens was more luminous than the total blended light. The dashed line is for no extinction to the lens, while the solid line is for the assumption that the lens is behind the same extinction as the source. Negative blending samples are shown artificially at blend light=25mag.
    }
    \label{fig:blend_lens_dlc}
\end{figure}

\begin{figure}
    \centering
    \includegraphics[width=9cm]{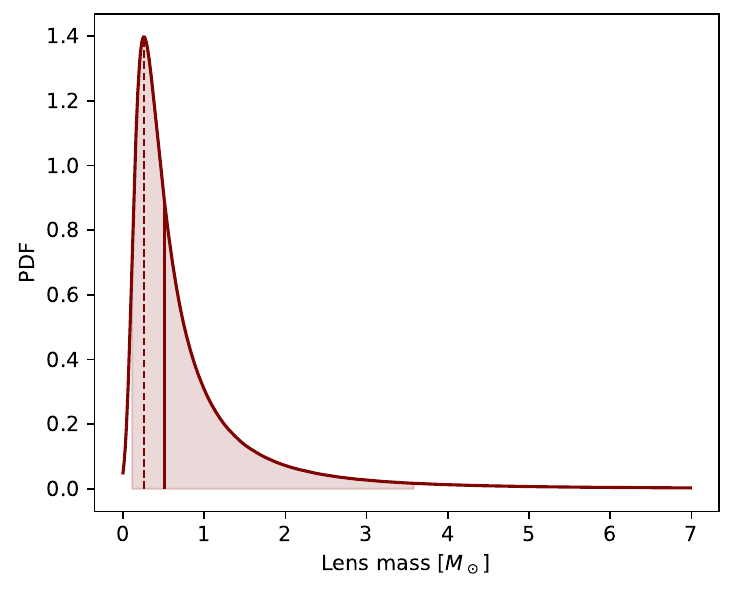}
    \caption{
    Probability density plot for the mass of the lens for G+F solution. The solid line marks the median and the dashed line marks the mode. The filled red area represents the $95\%$ confidence interval.}
    \label{fig:mass_hist}
\end{figure}

\begin{figure}
    \centering
    \includegraphics[width=9cm]{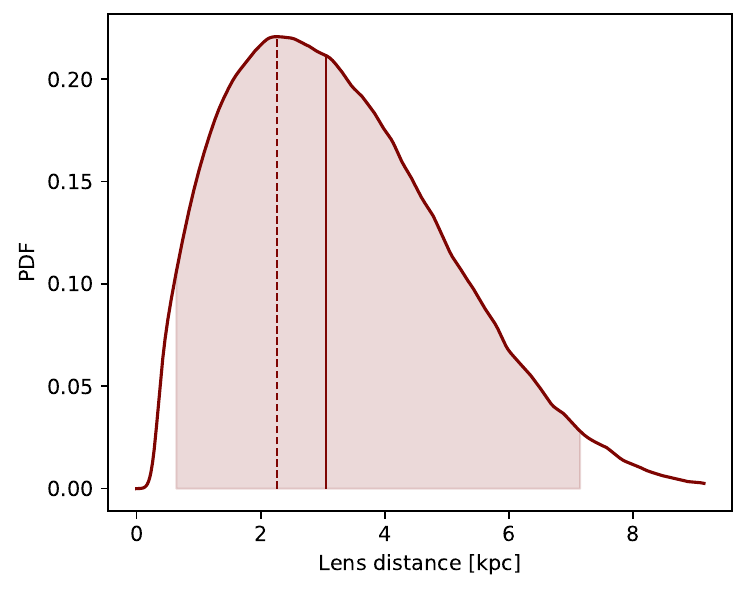}
    \caption{
     Probability density plot for distance to the lens for G+F solution. The solid line marks the median, and the dashed line marks the mode. The filled red area represents the $95\%$ confidence interval.}
    \label{fig:distance_hist}
\end{figure}

The results of this analysis are shown in Figures \ref{fig:mass_distance_dlc} and \ref{fig:blend_lens_dlc}. 
The histograms of the distribution of the lens mass and lens distance are visible in Figures \ref{fig:mass_hist} and \ref{fig:distance_hist}.
Table \ref{tab:lensScenarios} contains the summary of the median values of the mass, distance, blend light, and lens light in the case of an MS star lens. 
For {\gaia}-only data model the median mass is
$\ML=0.50^{+3.01}_{-0.40}~M_\odot$
and distance 
$\DL= 3.08^{4.09}_{-2.45}$ kpc.
For combined {\gaia} and follow-up data, the median mass and distance are 
$\ML = 0.51^{+3.07}_{-0.40}~M_\odot$ and  
$\DL = 3.05^{+4.10}_{-2.42}$ kpc, respectively. 
Modes of the distributions are, respectively, 
$\ML= 0.27~M_\odot$, $\DL = 2.31$ kpc, for G model and 
$\ML= 0.28~M_\odot$, $\DL = 2.26$ kpc, for G+F model.

Figure \ref{fig:blend_lens_dlc} contains the comparison of the light of the blend obtained from the microlensing model and the light of the lens if the lens is the MS star. 
Lines divide the plot area into two cases: above both lines prevail the scenario where the MS is justified given the blending. 
Below the lines, the light of the lens as an MS star is greater than the actual light of the lens we get from the microlensing model, suggesting a dark lens scenario.
The solid line denotes the scenario, when the value of the extinction is equal to the one for the source, while the dashed line assumes no extinction to the lens at all.
The luminous lens dark-lens scenario is preferred with 57\% to 63\% probability (for {\gaia}-only solution) and 58\% to 64\% for G+F model, with the range of probabilities resulting from a range of possible extinction values to the lens.




\begin{table}
        \centering
        \caption{
        \label{tab:lensScenarios}
             Lens masses $\ML$, distances $\DL$ and size of the Einstein Radius $\theta_E$ for the microlensing solutions.}
        \begin{tabular}{c c c}
            \hline
            \noalign{\smallskip}
            Parameter & G & G+F \\
            \noalign{\smallskip}
            \hline
            \hline
            \noalign{\smallskip}
            $G_{bl}$~[mag] & $ND$ & $>22.6$ \\ 
            \noalign{\smallskip}
             \hline
            \noalign{\smallskip}
            \hline
            \noalign{\smallskip}
            $\ML$~[$M_\odot$] & $0.50^{+3.01}_{-0.40}$ &  $0.51^{+3.07}_{-0.40}$ \\
            \noalign{\smallskip}
            $\DL$~[kpc] & $3.08^{+4.09}_{-2.45}$ &  $3.05^{+4.10}_{-2.42}$ \\
            \noalign{\smallskip}
            $\theta_E$~[mas] & $0.87^{+5.35}_{-0.68}$ &  $0.90^{+5.37}_{-0.70}$ \\
            \noalign{\smallskip}
            \noalign{\smallskip}
            Prob(DL) &  57.2\%-63.4\% &  58.4\%-64.5\% \\
            \noalign{\smallskip}
            $\mathrm{SpT}_\mathrm{MS}$ & M1V & M1V  \\
            \noalign{\smallskip}
            $G_\mathrm{MS}$[mag] & 22.1-21.3 & 22.1-21.3 \\
            \hline
            \noalign{\smallskip}
        \end{tabular}
        \tablefoot{$G_{bl}$ is the limit for the brightness of the lens computed using $f_S$ parameter for $\Gaia$ data, note that for $f_S>1$ the blend magnitude can not be determined (ND).
        $SpT_\mathrm{MS}$ is the spectral type of the lens if it was a main sequence star at $\ML$.
        $G_\mathrm{MS}$ is the brightness of the lens, with and without extinction, if it was a main sequence star of mass similar to $\ML$ located at the median distance $\DL$. 
        The absolute magnitude in the $G$-band has been taken from \cite{PecautMamajek2013}. Prob(DL) is the probability of the lens being a dark remnant with and without extinction}
\end{table}

%
\section{Discussion}



The microlensing event Gaia19dke lasted for about 2000 days (more than 5 years), making it one of the longest events ever studied. The annual parallax due to Earth's orbital motion caused a very strong microlensing parallax anomaly to a standard Paczynski curve visible in the light curve as a series of multiple peaks\citep{Smith2002}. Typically in parallax events, the $u_0$ sign and $\pi_E$ (N-E) degeneracy are present, in particular, this is common in the case of Bulge events, where the Ecliptic crosses the Galactic Plane. In Gaia19dke, located at about 300,50 deg in ecliptic coordinates, there was only one unique solution found for microlensing parallax. The annual parallax is measurable from \Gaia data alone. However, adding the extensive ground-based follow-up observations improves the parallax vector measurement by a factor of 3. The space-based parallax between \Gaia and the ground-based observatories was included in the model, however, was too small to be detected.

The microlensing parameters measured based solely on photometric \Gaia data were derived with an accuracy of about 1-3\%.
The addition of extensive ground-based follow-up observations improved the determination of all parameters by a factor of about 3. In particular, the improvement is the strongest in the case of the impact parameter $u_0$ and Einstein time-scale $t_E$, while the uncertainty on the parallax vector is about 0.8\% with the follow-up data. 

More importantly, the blending parameter for $\Gaia$ data has been determined more accurately when including follow-up data, from 4\% to around 1\%, which additionally supports the dark-lens case scenario. 
Blending parameters determined for all other modelled bands additionally confirm there is no or very little extra light apart from the source, with values of the blending very close to 1. Combining this information with no detection of any additional sources in the high-resolution image from 'Alopeke, strengthens the dark or very faint lens case. 
We decided to use $\Gaia$'s blending parameter in the lens nature determination in Section \ref{sec:lens} because GSA data covers both sides of the light curve, both its rising and declining parts as well as the baseline before the event, while other data sets covered only the central part of the event.

Microlensing in Gaia19dke allows us to determine the lens mass and its distance only because we measure the microlensing parallax and we use the priors on the lens proper motions as well as its distance and slope of the mass function. The results for mass and distance of the lens are summarised in Table \ref{tab:lensScenarios}, however, it should be noted that all the resulting posterior distributions are non-symmetric. Nevertheless, when using median values for mass and distance for either solution, we find the lens would need to be an M1V spectral-type star if it was a main sequence object. Placed at a median distance it would shine at 21.3 mag or 22.1 mag if all extinction measured to the source was in front of the lens. When compared with the amount of blending we measure in the light curve and its microlensing model, we can rule out such a scenario of a luminous lens. For a more massive lens, its distance would be even shorter, yielding an increase in the brightness of the alleged main sequence star. Only masses lower than the median would be possible to be explained within the observed bounds for blended light. The total integral over the parameter space yields between 57 and 64\% dark lens probability for both G and G+F models, the range resulting from including none or all extinction to the lens light.

 The high angular-resolution image obtained on 2020 Aug.9 with ’Alopeke instrument at the Gemini telescope does not show any visible additional object within 20 mas. From the long-term microlensing light curve analysis which started on the 8th of August 2019 and involved a massive ground telescope follow-up campaign that allowed us to collect a very detailed light curve for Gaia19dke, we also did not detect any binary lens signatures, typically visible as deviations to standard lensing curve and sharp caustic crossings. This strengthens the explanation of the shape of the light curve as microlensing by a single lens, affected by the parallax effect due to the Earth’s orbit. We, therefore, suggest the lensing event could have been caused by a stellar remnant.

Stellar evolution theory predicts that White Dwarfs (WDs) are the most common stellar remnants in the Galaxy. 
However, it is important to notice that, because of low brightness, the detection of WD is challenging. The majority of known WDs were found within the around 100~pc \cite{Fusillo2019MNRAS.482.4570G}, consequently, a full understanding of the WD population is far from complete. According to \cite{Takahashi2013ApJ...771...28T} the upper mass limit for a WD is 1.367~$M_\odot$, confirmed with the recent discovery of 1.35 $M_\odot$ WD \citep{Caiazzo2021Natur}. The most common mass of WD, however, falls within the range of $0.6 M_\odot$ - $0.7 M_\odot$ \citep{McCleery2020MNRAS.499.1890M}.

The most probable mass of the lens in our models is around 0.5 $M_\odot$, making the WD option most feasible. 
However, the possible mass range for the lens (Fig. \ref{fig:mass_hist}) also spans to larger masses, hence we can not rule out even a nearby neutron star scenario. 


Gaia19dke event is an excellent example of microlensing events for which $\Gaia$'s astrometric time-series will provide an actual measurement of the lens mass and distance through measurement of a tiny displacement of the source star due to microlensing \citep{Dominik2000,BelokurovEvans2002}. In the case of non-blended events like this one, the shift in the position of the source is of the order of the size of the Einstein Radius. Using Galaxy priors we estimate this size to be about 1 mas, hence easily detectable in the $\Gaia$ astrometric data \citep{Rybicki2018, Jablonska2022, Wyrzykowski2023}. 

\section{Conclusions}

In this work, we presented the investigation and analysis of a very long multi-peak microlensing event Gaia19dke located in the Galactic Disk, discovered by the {\gaia} space satellite. The event exhibited a microlensing parallax effect perturbed by the Earth's orbital motion. The investigation is based on {\gaia} data and ground follow-up photometry and spectroscopy follow--up observations. We determined the source star distance to $D_S = (4.9 \pm 1.2)$~kpc and we estimated the lens mass of $M_L = (0.50^{+3.07}_{-0.40}) M_\odot$ and its distance of $D_L = (3.05 ^{+4.10}_{-2.42})$~kpc for the model including both $\Gaia$ and ground-based data. Since essentially all of the detected light is coming from the source, a possible explanation is that the lens is a dark remnant candidate, most likely a single WD star, but a neutron star can also be considered.

The conclusive answer to the question on the nature of the lens will come with the $\Gaia$ astrometric time-series data to be released within DR4 (part until mid-2019) and DR5 (all remaining data). Additionally, the high-resolution AO-assisted observations of the source star in about a decade should provide strong confirmation on the dark lens in case of a non-detection of the lens\citep[e.g.][]{Blackman2021Nature}.



\section*{Acknowledgments}
%
This work is supported by Polish NCN grants: Daina No. 2017/27/L/ST9/03221, grant No. S-LL-19-2 of the Research Council of Lithuania, Harmonia No. 2018/30/M/ST9/00311, Preludium No. 2017/25/N/ST9/01253,
Opus No. 2017/25/B/ST9/02805 and MNiSW grant DIR/WK/2018/12. 
This project used data obtained via BHTOM (https://bhtom.space), which has received funding from the European Union's Horizon 2020 research and innovation program under grant agreements No. 730890 and 101004719.
We thank LT Support Astronomers for their help with observations and data reduction. HHE also thanks TUBITAK National Observatory for partial support in using the T100 telescope with project number 21AT100-1799 (and our sincere thanks to the whole of humanity that came to the aid of the earthquake disaster in T\"urkiye). Observations were carried out under OPTICON programmes XOL19B040 (PI: P.~Zieli{\'n}ski). The Liverpool Telescope is operated on the island of La Palma by Liverpool John Moores University in the Spanish Observatorio del Roque de Los Muchachos of the Instituto de Astrofisica de Canarias with financial support from the UK Science and Technology Facilities Council. We acknowledge ESA {\gaia}, DPAC and the Photometric Science Alerts Team (\url{http://gsaweb.ast.cam.ac.uk/alerts}).
This paper made use of the Whole Sky Database (wsdb) created by Sergey Koposov and maintained at the Institute of Astronomy, Cambridge by Sergey Koposov, Vasily Belokurov and Wyn Evans with financial support from the Science \& Technology Facilities Council (STFC) and the European Research Council (ERC), with the use of the Q3C software 
(\url{http://adsabs.harvard.edu/abs/2006ASPC..351..735K}).
The LBT is an international collaboration among institutions in the United States, Italy and Germany. LBT Corporation partners are The University of Arizona on behalf of the Arizona university system; Istituto Nazionale di Astrofisica, Italy; LBT Beteiligungsgesellschaft, Germany, representing the Max-Planck Society, the Astrophysical Institute Potsdam, and Heidelberg University; The Ohio State University, and The Research Corporation, on behalf of The University of Notre Dame, University of Minnesota and the University of Virginia.
Some of the observations in the paper made use of the High-Resolution Imaging instrument ‘Alopeke obtained under Gemini LLP Proposal Number: GN/S-2021A-LP-105. ‘Alopeke was funded by the NASA Exoplanet Exploration Program and built at the NASA Ames Research Center by Steve B. Howell, Nic Scott, Elliott P. Horch, and Emmett Quigley. Alopeke was mounted on the Gemini North (and/or South) telescope of the international Gemini Observatory, a program of NSF’s OIR Lab, which is managed by the Association of Universities for Research in Astronomy (AURA) under a cooperative agreement with the National Science Foundation. on behalf of the Gemini partnership: the National Science Foundation (United States), National Research Council (Canada), Agencia Nacional de Investigación y Desarrollo (Chile), Ministerio de Ciencia, Tecnologia e Innovacion (Argentina), Ministerio da Ciencia, Tecnologia, Inovacoes e Comunicacoes (Brazil), and Korea Astronomy and Space Science Institute (Republic of Korea). T.G. was supported by the Scientific Research Projects Coordination Unit of Istanbul University, project number: FBG-2017-23943 and the Turkish Republic, Presidency of Strategy and Budget project, project number: 2016K121370.”
G. Damljanovi{\'c} and M. Stojanovi{\'c} acknowledge support by the Astronomical station Vidojevica, funding from the Ministry of Science, Technological Development and Innovation of the Republic of Serbia (contract No. 451-03-47/2023-01/200002), by the EC through project BELISSIMA (call FP7-REGPOT-2010-5, No. 265772), the observing and financial grant support from the Institute of Astronomy and Rozhen NAO BAS through the bilateral SANU-BAN joint research project "GAIA astrometry and fast variable astronomical objects", and support by the SANU project F-187.
Adam Popowicz was responsible for automation and running remote observations at Otivar observatory and was supported by grant BK-236/RAu-11/2023.
YT acknowledges the support of the DFG priority program SPP 1992 “Exploring the Diversity of Extrasolar Planets” (TS 356/3-1).
Josep Manel Carrasco was (partially) supported by the Spanish MICIN/AEI/10.13039/501100011033 and by "ERDF A way of making Europe" by the “European Union” through grant PID2021-122842OB-C21, and the Institute of Cosmos Sciences University of Barcelona (ICCUB, Unidad de Excelencia ’Mar\'{\i}a de Maeztu’) through grant CEX2019-000918-M. The Joan Oró Telescope (TJO) of the Montsec Observatory (OdM) is owned by the Catalan Government and operated by the Institute for Space Studies of Catalonia (IEEC). This work was funded by ANID, Millennium Science Initiative, ICN12\_009.
Supachai Awiphan was supported by a National Astronomical Research Institute of Thailand (NARIT) and Thailand Science Research and Innovation (TSRI) research grant. Nawapon Nakharutai acknowledges the support of Chiang Mai University.
This research is partially supported by the Optical and Infrared
Synergetic Telescopes for Education and Research (OISTER) program
funded by the MEXT of Japan. AF is supported by JSPS KAKENHI Grant Number JP17H02871.
RFJ acknowledges funding by ANID's Millennium Science Initiative
through grant ICN12\textunderscore 009, awarded to the Millennium
Institute of Astrophysics (MAS), and by ANID's Basal project FB210003.

\bibliography{bibs}

\begin{thebibliography}{90}
\expandafter\ifx\csname natexlab\endcsname\relax\def\natexlab#1{#1}\fi

\bibitem[{{Bachelet} {et~al.}(2022){Bachelet}, {Zieli{\'n}ski}, {Gromadzki},
  {Gezer}, {Rybicki}, {Kruszy{\'n}ska}, {Ihanec}, {Wyrzykowski}, {Street},
  {Tsapras}, {Hundertmark}, {Cassan}, {Harbeck}, \& {Rabus}}]{Bachelet2022}
{Bachelet}, E., {Zieli{\'n}ski}, P., {Gromadzki}, M., {et~al.} 2022, \aap, 657,
  A17

\bibitem[{{Bailer-Jones} {et~al.}(2021){Bailer-Jones}, {Rybizki}, {Fouesneau},
  {Demleitner}, \& {Andrae}}]{2021Bailer-Jones}
{Bailer-Jones}, C.~A.~L., {Rybizki}, J., {Fouesneau}, M., {Demleitner}, M., \&
  {Andrae}, R. 2021, \aj, 161, 147

\bibitem[{{Bailyn} {et~al.}(1998){Bailyn}, {Jain}, {Coppi}, \&
  {Orosz}}]{Bailyn1998}
{Bailyn}, C.~D., {Jain}, R.~K., {Coppi}, P., \& {Orosz}, J.~A. 1998, \apj, 499,
  367

\bibitem[{{Bellm} {et~al.}(2019){Bellm}, {Kulkarni}, {Graham}, {Dekany},
  {Smith}, {Riddle}, {Masci}, {Helou}, {Prince}, {Adams}, {Barbarino},
  {Barlow}, {Bauer}, {Beck}, {Belicki}, {Biswas}, {Blagorodnova}, {Bodewits},
  {Bolin}, {Brinnel}, {Brooke}, {Bue}, {Bulla}, {Burruss}, {Cenko}, {Chang},
  {Connolly}, {Coughlin}, {Cromer}, {Cunningham}, {De}, {Delacroix}, {Desai},
  {Duev}, {Eadie}, {Farnham}, {Feeney}, {Feindt}, {Flynn}, {Franckowiak},
  {Frederick}, {Fremling}, {Gal-Yam}, {Gezari}, {Giomi}, {Goldstein},
  {Golkhou}, {Goobar}, {Groom}, {Hacopians}, {Hale}, {Henning}, {Ho}, {Hover},
  {Howell}, {Hung}, {Huppenkothen}, {Imel}, {Ip}, {Ivezi{\'c}}, {Jackson},
  {Jones}, {Juric}, {Kasliwal}, {Kaspi}, {Kaye}, {Kelley}, {Kowalski},
  {Kramer}, {Kupfer}, {Landry}, {Laher}, {Lee}, {Lin}, {Lin}, {Lunnan},
  {Giomi}, {Mahabal}, {Mao}, {Miller}, {Monkewitz}, {Murphy}, {Ngeow},
  {Nordin}, {Nugent}, {Ofek}, {Patterson}, {Penprase}, {Porter}, {Rauch},
  {Rebbapragada}, {Reiley}, {Rigault}, {Rodriguez}, {van Roestel}, {Rusholme},
  {van Santen}, {Schulze}, {Shupe}, {Singer}, {Soumagnac}, {Stein}, {Surace},
  {Sollerman}, {Szkody}, {Taddia}, {Terek}, {Van Sistine}, {van Velzen},
  {Vestrand}, {Walters}, {Ward}, {Ye}, {Yu}, {Yan}, \& {Zolkower}}]{2019ZTF}
{Bellm}, E.~C., {Kulkarni}, S.~R., {Graham}, M.~J., {et~al.} 2019, \pasp, 131,
  018002

\bibitem[{{Belokurov} \& {Evans}(2002)}]{BelokurovEvans2002}
{Belokurov}, V.~A. \& {Evans}, N.~W. 2002, \mnras, 331, 649

\bibitem[{{Bird} {et~al.}(2016){Bird}, {Cholis}, {Mu{\~n}oz},
  {Ali-Ha{\"i}moud}, {Kamionkowski}, {Kovetz}, {Raccanelli}, \&
  {Riess}}]{Bird2016}
{Bird}, S., {Cholis}, I., {Mu{\~n}oz}, J.~B., {et~al.} 2016, Physical Review
  Letters, 116, 201301

\bibitem[{{Blackman} {et~al.}(2021){Blackman}, {Beaulieu}, {Bennett},
  {Danielski}, {Alard}, {Cole}, {Vandorou}, {Ranc}, {Terry}, {Bhattacharya},
  {Bond}, {Bachelet}, {Veras}, {Koshimoto}, {Batista}, \&
  {Marquette}}]{Blackman2021Nature}
{Blackman}, J.~W., {Beaulieu}, J.~P., {Bennett}, D.~P., {et~al.} 2021, \nat,
  598, 272

\bibitem[{{Blanco-Cuaresma}(2019)}]{BlancoCuaresma2019}
{Blanco-Cuaresma}, S. 2019, \mnras, 486, 2075

\bibitem[{{Blanco-Cuaresma} {et~al.}(2014){Blanco-Cuaresma}, {Soubiran},
  {Heiter}, \& {Jofr{\'e}}}]{BlancoCuaresma2014}
{Blanco-Cuaresma}, S., {Soubiran}, C., {Heiter}, U., \& {Jofr{\'e}}, P. 2014,
  \aap, 569, A111

\bibitem[{{Bressan} {et~al.}(2012){Bressan}, {Marigo}, {Girardi}, {Salasnich},
  {Dal Cero}, {Rubele}, \& {Nanni}}]{Bressan2012}
{Bressan}, A., {Marigo}, P., {Girardi}, L., {et~al.} 2012, \mnras, 427, 127

\bibitem[{{Brown} {et~al.}(2013{\natexlab{a}}){Brown}, {Baliber}, {Bianco},
  {Bowman}, {Burleson}, {Conway}, {Crellin}, {Depagne}, {De Vera}, {Dilday},
  {Dragomir}, {Dubberley}, {Eastman}, {Elphick}, {Falarski}, {Foale}, {Ford},
  {Fulton}, {Garza}, {Gomez}, {Graham}, {Greene}, {Haldeman}, {Hawkins},
  {Haworth}, {Haynes}, {Hidas}, {Hjelstrom}, {Howell}, {Hygelund}, {Lister},
  {Lobdill}, {Martinez}, {Mullins}, {Norbury}, {Parrent}, {Paulson}, {Petry},
  {Pickles}, {Posner}, {Rosing}, {Ross}, {Sand}, {Saunders}, {Shobbrook},
  {Shporer}, {Street}, {Thomas}, {Tsapras}, {Tufts}, {Valenti}, {Vander Horst},
  {Walker}, {White}, \& {Willis}}]{Brown2013}
{Brown}, T.~M., {Baliber}, N., {Bianco}, F.~B., {et~al.} 2013{\natexlab{a}},
  \pasp, 125, 1031

\bibitem[{{Brown} {et~al.}(2013{\natexlab{b}}){Brown}, {Baliber}, {Bianco},
  {Bowman}, {Burleson}, {Conway}, {Crellin}, {Depagne}, {De Vera}, {Dilday},
  {Dragomir}, {Dubberley}, {Eastman}, {Elphick}, {Falarski}, {Foale}, {Ford},
  {Fulton}, {Garza}, {Gomez}, {Graham}, {Greene}, {Haldeman}, {Hawkins},
  {Haworth}, {Haynes}, {Hidas}, {Hjelstrom}, {Howell}, {Hygelund}, {Lister},
  {Lobdill}, {Martinez}, {Mullins}, {Norbury}, {Parrent}, {Paulson}, {Petry},
  {Pickles}, {Posner}, {Rosing}, {Ross}, {Sand}, {Saunders}, {Shobbrook},
  {Shporer}, {Street}, {Thomas}, {Tsapras}, {Tufts}, {Valenti}, {Vander Horst},
  {Walker}, {White}, \& {Willis}}]{LCO}
{Brown}, T.~M., {Baliber}, N., {Bianco}, F.~B., {et~al.} 2013{\natexlab{b}},
  \pasp, 125, 1031

\bibitem[{{Caiazzo} {et~al.}(2021){Caiazzo}, {Burdge}, {Fuller}, {Heyl},
  {Kulkarni}, {Prince}, {Richer}, {Schwab}, {Andreoni}, {Bellm}, {Drake},
  {Duev}, {Graham}, {Helou}, {Mahabal}, {Masci}, {Smith}, \&
  {Soumagnac}}]{Caiazzo2021Natur}
{Caiazzo}, I., {Burdge}, K.~B., {Fuller}, J., {et~al.} 2021, \nat, 596, E15

\bibitem[{{Cardelli} {et~al.}(1989{\natexlab{a}}){Cardelli}, {Clayton}, \&
  {Mathis}}]{Cardelli1989}
{Cardelli}, J.~A., {Clayton}, G.~C., \& {Mathis}, J.~S. 1989{\natexlab{a}},
  \apj, 345, 245

\bibitem[{{Cardelli} {et~al.}(1989{\natexlab{b}}){Cardelli}, {Clayton}, \&
  {Mathis}}]{Cardelli}
{Cardelli}, J.~A., {Clayton}, G.~C., \& {Mathis}, J.~S. 1989{\natexlab{b}},
  \apj, 345, 245

\bibitem[{{Carr} \& {Silk}(2018)}]{Carr2018}
{Carr}, B. \& {Silk}, J. 2018, \mnras, 478, 3756

\bibitem[{{Cassan} {et~al.}(2022){Cassan}, {Ranc}, {Absil}, {Wyrzykowski},
  {Rybicki}, {Bachelet}, {Le Bouquin}, {Hundertmark}, {Street}, {Surdej},
  {Tsapras}, {Wambsganss}, \& {Wertz}}]{CassanVLTI}
{Cassan}, A., {Ranc}, C., {Absil}, O., {et~al.} 2022, Nature Astronomy, 6, 121

\bibitem[{{Clesse} \& {Garc{\'\i}a-Bellido}(2015)}]{Clesse2015}
{Clesse}, S. \& {Garc{\'\i}a-Bellido}, J. 2015, \prd, 92, 023524

\bibitem[{{Damljanovi{\'c}} {et~al.}(2023){Damljanovi{\'c}}, {Stojanovi{\'c}},
  {Bachev}, \& {Boeva}}]{ASV}
{Damljanovi{\'c}}, G., {Stojanovi{\'c}}, M., {Bachev}, R., \& {Boeva}, S. 2023,
  in Proceedings of the XIII Bulgarian-Serbian Astronomical Conference,
  Vol.~25, 43--51

\bibitem[{{Dominik} \& {Sahu}(2000)}]{Dominik2000}
{Dominik}, M. \& {Sahu}, K.~C. 2000, \apj, 534, 213

\bibitem[{{Dong} {et~al.}(2019){Dong}, {M{\'e}rand}, {Delplancke-Str{\"o}bele},
  {Gould}, {Chen}, {Post}, {Kochanek}, {Stanek}, {Christie}, {Mutel},
  {Natusch}, {Holoien}, {Prieto}, {Shappee}, \& {Thompson}}]{2019DongVLTI}
{Dong}, S., {M{\'e}rand}, A., {Delplancke-Str{\"o}bele}, F., {et~al.} 2019,
  \apj, 871, 70

\bibitem[{{Dutra} {et~al.}(2002){Dutra}, {Santiago}, \& {Bica}}]{Dutra}
{Dutra}, C.~M., {Santiago}, B.~X., \& {Bica}, E. 2002, \aap, 381, 219

\bibitem[{{Farr} {et~al.}(2011){Farr}, {Sravan}, {Cantrell}, {Kreidberg},
  {Bailyn}, {Mandel}, \& {Kalogera}}]{Farr:2010}
{Farr}, W.~M., {Sravan}, N., {Cantrell}, A., {et~al.} 2011, \apj, 741, 103

\bibitem[{Foreman-Mackey(2016)}]{corner}
Foreman-Mackey, D. 2016, The Journal of Open Source Software, 1, 24

\bibitem[{{Foreman-Mackey} {et~al.}(2013){Foreman-Mackey}, {Hogg}, {Lang}, \&
  {Goodman}}]{EMCEE}
{Foreman-Mackey}, D., {Hogg}, D.~W., {Lang}, D., \& {Goodman}, J. 2013, \pasp,
  125, 306

\bibitem[{{Gaia Collaboration}(2020)}]{GaiaEDR3Cat}
{Gaia Collaboration}. 2020, VizieR Online Data Catalog, I/350

\bibitem[{{Gaia Collaboration} {et~al.}(2018){Gaia Collaboration}, {Brown},
  {Vallenari}, {Prusti}, {de Bruijne}, {Babusiaux}, {Bailer-Jones}, {Biermann},
  {Evans}, {Eyer}, \& et~al.}]{GaiaDR2}
{Gaia Collaboration}, {Brown}, A.~G.~A., {Vallenari}, A., {et~al.} 2018, \aap,
  616, A1

\bibitem[{{Gaia Collaboration} {et~al.}(2023{\natexlab{a}}){Gaia
  Collaboration}, {Montegriffo}, {Bellazzini}, {De Angeli}, {Andrae},
  {Barstow}, {Bossini}, {Bragaglia}, {Burgess}, {Cacciari}, {Carrasco},
  {Chornay}, {Delchambre}, {Evans}, {Fouesneau}, {Fr{\'e}mat}, {Garabato},
  {Jordi}, {Manteiga}, {Massari}, {Palaversa}, {Pancino}, {Riello}, {Ruz
  Mieres}, {Sanna}, {Santove{\~n}a}, {Sordo}, {Vallenari}, {Walton}, {Brown},
  {Prusti}, {de Bruijne}, {Arenou}, {Babusiaux}, {Biermann}, {Creevey},
  {Ducourant}, {Eyer}, {Guerra}, {Hutton}, {Klioner}, {Lammers}, {Lindegren},
  {Luri}, {Mignard}, {Panem}, {Pourbaix}, {Randich}, {Sartoretti}, {Soubiran},
  {Tanga}, {Bailer-Jones}, {Bastian}, {Drimmel}, {Jansen}, {Katz}, {Lattanzi},
  {van Leeuwen}, {Bakker}, {Casta{\~n}eda}, {Fabricius}, {Galluccio},
  {Guerrier}, {Heiter}, {Masana}, {Messineo}, {Mowlavi}, {Nicolas},
  {Nienartowicz}, {Pailler}, {Panuzzo}, {Riclet}, {Roux}, {Seabroke},
  {Th{\'e}venin}, {Gracia-Abril}, {Portell}, {Teyssier}, {Altmann}, {Audard},
  {Bellas-Velidis}, {Benson}, {Berthier}, {Blomme}, {Busonero}, {Busso},
  {C{\'a}novas}, {Carry}, {Cellino}, {Cheek}, {Clementini}, {Damerdji},
  {Davidson}, {de Teodoro}, {Nu{\~n}ez Campos}, {Dell'Oro}, {Esquej},
  {Fern{\'a}ndez-Hern{\'a}ndez}, {Fraile}, {Garc{\'\i}a-Lario}, {Gosset},
  {Haigron}, {Halbwachs}, {Hambly}, {Harrison}, {Hern{\'a}ndez}, {Hestroffer},
  {Hodgkin}, {Holl}, {Jan{\ss}en}, {Jevardat de Fombelle}, {Jordan},
  {Krone-Martins}, {Lanzafame}, {L{\"o}ffler}, {Marchal}, {Marrese},
  {Moitinho}, {Muinonen}, {Osborne}, {Pauwels}, {Recio-Blanco}, {Reyl{\'e}},
  {Rimoldini}, {Roegiers}, {Rybizki}, {Sarro}, {Siopis}, {Smith}, {Sozzetti},
  {Utrilla}, {van Leeuwen}, {Abbas}, {{\'A}brah{\'a}m}, {Abreu Aramburu},
  {Aerts}, {Aguado}, {Ajaj}, {Aldea-Montero}, {Altavilla}, {{\'A}lvarez},
  {Alves}, {Anderson}, {Anglada Varela}, {Antoja}, {Baines}, {Baker},
  {Balaguer-N{\'u}{\~n}ez}, {Balbinot}, {Balog}, {Barache}, {Barbato},
  {Barros}, {Bartolom{\'e}}, {Bassilana}, {Bauchet}, {Becciani}, {Berihuete},
  {Bernet}, {Bertone}, {Bianchi}, {Binnenfeld}, {Blanco-Cuaresma}, {Boch},
  {Bombrun}, {Bouquillon}, {Bramante}, {Breedt}, {Bressan}, {Brouillet},
  {Brugaletta}, {Bucciarelli}, {Burlacu}, {Butkevich}, {Buzzi}, {Caffau},
  {Cancelliere}, {Cantat-Gaudin}, {Carballo}, {Carlucci}, {Carnerero},
  {Casamiquela}, {Castellani}, {Castro-Ginard}, {Chaoul}, {Charlot}, {Chemin},
  {Chiaramida}, {Chiavassa}, {Comoretto}, {Contursi}, {Cooper}, {Cornez},
  {Cowell}, {Crifo}, {Cropper}, {Crosta}, {Crowley}, {Dafonte}, {Dapergolas},
  {David}, {de Laverny}, {De Luise}, {De March}, {De Ridder}, {de Souza}, {de
  Torres}, {del Peloso}, {del Pozo}, {Delbo}, {Delgado}, {Delisle}, {Demouchy},
  {Dharmawardena}, {Diakite}, {Diener}, {Distefano}, {Dolding}, {Enke},
  {Fabre}, {Fabrizio}, {Faigler}, {Fedorets}, {Fernique}, {Figueras},
  {Fournier}, {Fouron}, {Fragkoudi}, {Gai}, {Garcia-Gutierrez},
  {Garcia-Reinaldos}, {Garc{\'\i}a-Torres}, {Garofalo}, {Gavel}, {Gavras},
  {Gerlach}, {Geyer}, {Giacobbe}, {Gilmore}, {Girona}, {Giuffrida}, {Gomel},
  {Gomez}, {Gonz{\'a}lez-N{\'u}{\~n}ez}, {Gonz{\'a}lez-Santamar{\'\i}a},
  {Gonz{\'a}lez-Vidal}, {Granvik}, {Guillout}, {Guiraud},
  {Guti{\'e}rrez-S{\'a}nchez}, {Guy}, {Hatzidimitriou}, {Hauser}, {Haywood},
  {Helmer}, {Helmi}, {Sarmiento}, {Hidalgo}, {H{\l}adczuk}, {Hobbs}, {Holland},
  {Huckle}, {Jardine}, {Jasniewicz}, {Jean-Antoine Piccolo},
  {Jim{\'e}nez-Arranz}, {Juaristi Campillo}, {Julbe}, {Karbevska}, {Kervella},
  {Khanna}, {Kordopatis}, {Korn}, {K{\'o}sp{\'a}l}, {Kostrzewa-Rutkowska},
  {Kruszy{\'n}ska}, {Kun}, {Laizeau}, {Lambert}, {Lanza}, {Lasne}, {Le
  Campion}, {Lebreton}, {Lebzelter}, {Leccia}, {Leclerc}, {Lecoeur-Taibi},
  {Liao}, {Licata}, {Lindstr{\'o}m}, {Lister}, {Livanou}, {Lobel}, {Lorca},
  {Loup}, {Madrero Pardo}, {Magdaleno Romeo}, {Managau}, {Mann}, {Marchant},
  {Marconi}, {Marcos}, {Marcos Santos}, {Mar{\'\i}n Pina}, {Marinoni},
  {Marocco}, {Marshall}, {Martin Polo}, {Mart{\'\i}n-Fleitas}, {Marton},
  {Mary}, {Masip}, {Mastrobuono-Battisti}, {Mazeh}, {McMillan}, {Messina},
  {Michalik}, {Millar}, {Mints}, {Molina}, {Molinaro}, {Moln{\'a}r}, {Monari},
  {Mongui{\'o}}, {Montero}, {Mor}, {Mora}, {Morbidelli}, {Morel}, {Morris},
  {Muraveva}, {Murphy}, {Musella}, {Nagy}, {Noval}, {Oca{\~n}a}, {Ogden},
  {Ordenovic}, {Osinde}, {Pagani}, {Pagano}, {Palicio}, {Pallas-Quintela},
  {Panahi}, {Payne-Wardenaar}, {Pe{\~n}alosa Esteller}, {Penttil{\"a}},
  {Pichon}, {Piersimoni}, {Pineau}, {Plachy}, {Plum}, {Poggio}, {Pr{\v{s}}a},
  {Pulone}, {Racero}, {Ragaini}, {Rainer}, {Raiteri}, {Ramos}, {Ramos-Lerate},
  {Re Fiorentin}, {Regibo}, {Richards}, {Rios Diaz}, {Ripepi}, {Riva}, {Rix},
  {Rixon}, {Robichon}, {Robin}, {Robin}, {Roelens}, {Rogues}, {Rohrbasser},
  {Romero-G{\'o}mez}, {Rowell}, {Royer}, {Rybicki}, {Sadowski}, {S{\'a}ez
  N{\'u}{\~n}ez}, {Sagrist{\`a} Sell{\'e}s}, {Sahlmann}, {Salguero}, {Samaras},
  {Sanchez Gimenez}, {Sarasso}, {Schultheis}, {Sciacca}, {Segol}, {Segovia},
  {S{\'e}gransan}, {Semeux}, {Shahaf}, {Siddiqui}, {Siebert}, {Siltala},
  {Silvelo}, {Slezak}, {Slezak}, {Smart}, {Snaith}, {Solano}, {Solitro},
  {Souami}, {Souchay}, {Spagna}, {Spina}, {Spoto}, {Steele},
  {Steidelm{\"u}ller}, {Stephenson}, {S{\"u}veges}, {Surdej}, {Szabados},
  {Szegedi-Elek}, {Taris}, {Taylor}, {Teixeira}, {Tolomei}, {Tonello}, {Torra},
  {Torra}, {Torralba Elipe}, {Trabucchi}, {Tsounis}, {Turon}, {Ulla}, {Unger},
  {Vaillant}, {van Dillen}, {van Reeven}, {Vanel}, {Vecchiato}, {Viala},
  {Vicente}, {Voutsinas}, {Wevers}, {Wyrzykowski}, {Yoldas}, {Yvard}, {Zhao},
  {Zorec}, {Zucker}, \& {Zwitter}}]{GaiaSP2023}
{Gaia Collaboration}, {Montegriffo}, P., {Bellazzini}, M., {et~al.}
  2023{\natexlab{a}}, \aap, 674, A33

\bibitem[{{Gaia Collaboration} {et~al.}(2016){Gaia Collaboration}, {Prusti},
  {de Bruijne}, {Brown}, {Vallenari}, {Babusiaux}, {Bailer-Jones}, {Bastian},
  {Biermann}, {Evans}, \& et~al.}]{Gaia}
{Gaia Collaboration}, {Prusti}, T., {de Bruijne}, J.~H.~J., {et~al.} 2016,
  \aap, 595, A1

\bibitem[{{Gaia Collaboration} {et~al.}(2023{\natexlab{b}}){Gaia
  Collaboration}, {Vallenari}, {Brown}, {Prusti}, {de Bruijne}, {Arenou},
  {Babusiaux}, {Biermann}, {Creevey}, {Ducourant}, {Evans}, {Eyer}, {Guerra},
  {Hutton}, {Jordi}, {Klioner}, {Lammers}, {Lindegren}, {Luri}, {Mignard},
  {Panem}, {Pourbaix}, {Randich}, {Sartoretti}, {Soubiran}, {Tanga}, {Walton},
  {Bailer-Jones}, {Bastian}, {Drimmel}, {Jansen}, {Katz}, {Lattanzi}, {van
  Leeuwen}, {Bakker}, {Cacciari}, {Casta{\~n}eda}, {De Angeli}, {Fabricius},
  {Fouesneau}, {Fr{\'e}mat}, {Galluccio}, {Guerrier}, {Heiter}, {Masana},
  {Messineo}, {Mowlavi}, {Nicolas}, {Nienartowicz}, {Pailler}, {Panuzzo},
  {Riclet}, {Roux}, {Seabroke}, {Sordo}, {Th{\'e}venin}, {Gracia-Abril},
  {Portell}, {Teyssier}, {Altmann}, {Andrae}, {Audard}, {Bellas-Velidis},
  {Benson}, {Berthier}, {Blomme}, {Burgess}, {Busonero}, {Busso},
  {C{\'a}novas}, {Carry}, {Cellino}, {Cheek}, {Clementini}, {Damerdji},
  {Davidson}, {de Teodoro}, {Nu{\~n}ez Campos}, {Delchambre}, {Dell'Oro},
  {Esquej}, {Fern{\'a}ndez-Hern{\'a}ndez}, {Fraile}, {Garabato},
  {Garc{\'\i}a-Lario}, {Gosset}, {Haigron}, {Halbwachs}, {Hambly}, {Harrison},
  {Hern{\'a}ndez}, {Hestroffer}, {Hodgkin}, {Holl}, {Jan{\ss}en}, {Jevardat de
  Fombelle}, {Jordan}, {Krone-Martins}, {Lanzafame}, {L{\"o}ffler}, {Marchal},
  {Marrese}, {Moitinho}, {Muinonen}, {Osborne}, {Pancino}, {Pauwels},
  {Recio-Blanco}, {Reyl{\'e}}, {Riello}, {Rimoldini}, {Roegiers}, {Rybizki},
  {Sarro}, {Siopis}, {Smith}, {Sozzetti}, {Utrilla}, {van Leeuwen}, {Abbas},
  {{\'A}brah{\'a}m}, {Abreu Aramburu}, {Aerts}, {Aguado}, {Ajaj},
  {Aldea-Montero}, {Altavilla}, {{\'A}lvarez}, {Alves}, {Anders}, {Anderson},
  {Anglada Varela}, {Antoja}, {Baines}, {Baker}, {Balaguer-N{\'u}{\~n}ez},
  {Balbinot}, {Balog}, {Barache}, {Barbato}, {Barros}, {Barstow},
  {Bartolom{\'e}}, {Bassilana}, {Bauchet}, {Becciani}, {Bellazzini},
  {Berihuete}, {Bernet}, {Bertone}, {Bianchi}, {Binnenfeld}, {Blanco-Cuaresma},
  {Blazere}, {Boch}, {Bombrun}, {Bossini}, {Bouquillon}, {Bragaglia},
  {Bramante}, {Breedt}, {Bressan}, {Brouillet}, {Brugaletta}, {Bucciarelli},
  {Burlacu}, {Butkevich}, {Buzzi}, {Caffau}, {Cancelliere}, {Cantat-Gaudin},
  {Carballo}, {Carlucci}, {Carnerero}, {Carrasco}, {Casamiquela}, {Castellani},
  {Castro-Ginard}, {Chaoul}, {Charlot}, {Chemin}, {Chiaramida}, {Chiavassa},
  {Chornay}, {Comoretto}, {Contursi}, {Cooper}, {Cornez}, {Cowell}, {Crifo},
  {Cropper}, {Crosta}, {Crowley}, {Dafonte}, {Dapergolas}, {David}, {David},
  {de Laverny}, {De Luise}, {De March}, {De Ridder}, {de Souza}, {de Torres},
  {del Peloso}, {del Pozo}, {Delbo}, {Delgado}, {Delisle}, {Demouchy},
  {Dharmawardena}, {Di Matteo}, {Diakite}, {Diener}, {Distefano}, {Dolding},
  {Edvardsson}, {Enke}, {Fabre}, {Fabrizio}, {Faigler}, {Fedorets}, {Fernique},
  {Fienga}, {Figueras}, {Fournier}, {Fouron}, {Fragkoudi}, {Gai},
  {Garcia-Gutierrez}, {Garcia-Reinaldos}, {Garc{\'\i}a-Torres}, {Garofalo},
  {Gavel}, {Gavras}, {Gerlach}, {Geyer}, {Giacobbe}, {Gilmore}, {Girona},
  {Giuffrida}, {Gomel}, {Gomez}, {Gonz{\'a}lez-N{\'u}{\~n}ez},
  {Gonz{\'a}lez-Santamar{\'\i}a}, {Gonz{\'a}lez-Vidal}, {Granvik}, {Guillout},
  {Guiraud}, {Guti{\'e}rrez-S{\'a}nchez}, {Guy}, {Hatzidimitriou}, {Hauser},
  {Haywood}, {Helmer}, {Helmi}, {Sarmiento}, {Hidalgo}, {Hilger},
  {H{\l}adczuk}, {Hobbs}, {Holland}, {Huckle}, {Jardine}, {Jasniewicz},
  {Jean-Antoine Piccolo}, {Jim{\'e}nez-Arranz}, {Jorissen}, {Juaristi
  Campillo}, {Julbe}, {Karbevska}, {Kervella}, {Khanna}, {Kontizas},
  {Kordopatis}, {Korn}, {K{\'o}sp{\'a}l}, {Kostrzewa-Rutkowska},
  {Kruszy{\'n}ska}, {Kun}, {Laizeau}, {Lambert}, {Lanza}, {Lasne}, {Le
  Campion}, {Lebreton}, {Lebzelter}, {Leccia}, {Leclerc}, {Lecoeur-Taibi},
  {Liao}, {Licata}, {Lindstr{\o}m}, {Lister}, {Livanou}, {Lobel}, {Lorca},
  {Loup}, {Madrero Pardo}, {Magdaleno Romeo}, {Managau}, {Mann}, {Manteiga},
  {Marchant}, {Marconi}, {Marcos}, {Marcos Santos}, {Mar{\'\i}n Pina},
  {Marinoni}, {Marocco}, {Marshall}, {Martin Polo}, {Mart{\'\i}n-Fleitas},
  {Marton}, {Mary}, {Masip}, {Massari}, {Mastrobuono-Battisti}, {Mazeh},
  {McMillan}, {Messina}, {Michalik}, {Millar}, {Mints}, {Molina}, {Molinaro},
  {Moln{\'a}r}, {Monari}, {Mongui{\'o}}, {Montegriffo}, {Montero}, {Mor},
  {Mora}, {Morbidelli}, {Morel}, {Morris}, {Muraveva}, {Murphy}, {Musella},
  {Nagy}, {Noval}, {Oca{\~n}a}, {Ogden}, {Ordenovic}, {Osinde}, {Pagani},
  {Pagano}, {Palaversa}, {Palicio}, {Pallas-Quintela}, {Panahi},
  {Payne-Wardenaar}, {Pe{\~n}alosa Esteller}, {Penttil{\"a}}, {Pichon},
  {Piersimoni}, {Pineau}, {Plachy}, {Plum}, {Poggio}, {Pr{\v{s}}a}, {Pulone},
  {Racero}, {Ragaini}, {Rainer}, {Raiteri}, {Rambaux}, {Ramos}, {Ramos-Lerate},
  {Re Fiorentin}, {Regibo}, {Richards}, {Rios Diaz}, {Ripepi}, {Riva}, {Rix},
  {Rixon}, {Robichon}, {Robin}, {Robin}, {Roelens}, {Rogues}, {Rohrbasser},
  {Romero-G{\'o}mez}, {Rowell}, {Royer}, {Ruz Mieres}, {Rybicki}, {Sadowski},
  {S{\'a}ez N{\'u}{\~n}ez}, {Sagrist{\`a} Sell{\'e}s}, {Sahlmann}, {Salguero},
  {Samaras}, {Sanchez Gimenez}, {Sanna}, {Santove{\~n}a}, {Sarasso},
  {Schultheis}, {Sciacca}, {Segol}, {Segovia}, {S{\'e}gransan}, {Semeux},
  {Shahaf}, {Siddiqui}, {Siebert}, {Siltala}, {Silvelo}, {Slezak}, {Slezak},
  {Smart}, {Snaith}, {Solano}, {Solitro}, {Souami}, {Souchay}, {Spagna},
  {Spina}, {Spoto}, {Steele}, {Steidelm{\"u}ller}, {Stephenson}, {S{\"u}veges},
  {Surdej}, {Szabados}, {Szegedi-Elek}, {Taris}, {Taylor}, {Teixeira},
  {Tolomei}, {Tonello}, {Torra}, {Torra}, {Torralba Elipe}, {Trabucchi},
  {Tsounis}, {Turon}, {Ulla}, {Unger}, {Vaillant}, {van Dillen}, {van Reeven},
  {Vanel}, {Vecchiato}, {Viala}, {Vicente}, {Voutsinas}, {Weiler}, {Wevers},
  {Wyrzykowski}, {Yoldas}, {Yvard}, {Zhao}, {Zorec}, {Zucker}, \&
  {Zwitter}}]{GaiaDR3}
{Gaia Collaboration}, {Vallenari}, A., {Brown}, A.~G.~A., {et~al.}
  2023{\natexlab{b}}, \aap, 674, A1

\bibitem[{{Gentile Fusillo} {et~al.}(2019){Gentile Fusillo}, {Tremblay},
  {G{\"a}nsicke}, {Manser}, {Cunningham}, {Cukanovaite}, {Hollands}, {Marsh},
  {Raddi}, {Jordan}, {Toonen}, {Geier}, {Barstow}, \&
  {Cummings}}]{Fusillo2019MNRAS.482.4570G}
{Gentile Fusillo}, N.~P., {Tremblay}, P.-E., {G{\"a}nsicke}, B.~T., {et~al.}
  2019, \mnras, 482, 4570

\bibitem[{{Gould}(2000)}]{Gould2000b}
{Gould}, A. 2000, \apj, 542, 785

\bibitem[{{Gould}(2004)}]{Gould2004}
{Gould}, A. 2004, \apj, 606, 319

\bibitem[{{Gould} {et~al.}(2004){Gould}, {Bennett}, \& {Alves}}]{GouldMass2004}
{Gould}, A., {Bennett}, D.~P., \& {Alves}, D.~R. 2004, \apj, 614, 404

\bibitem[{{Gould} \& {Yee}(2014)}]{GouldYee2014}
{Gould}, A. \& {Yee}, J.~C. 2014, \apj, 784, 64

\bibitem[{{Grevesse} {et~al.}(2007){Grevesse}, {Asplund}, \&
  {Sauval}}]{Grevesse2007}
{Grevesse}, N., {Asplund}, M., \& {Sauval}, A.~J. 2007, \ssr, 130, 105

\bibitem[{{Gustafsson} {et~al.}(2008){Gustafsson}, {Edvardsson}, {Eriksson},
  {J{\o}rgensen}, {Nordlund}, \& {Plez}}]{Gustafsson2008}
{Gustafsson}, B., {Edvardsson}, B., {Eriksson}, K., {et~al.} 2008, \aap, 486,
  951

\bibitem[{{Henry} {et~al.}(2000){Henry}, {Fekel}, {Henry}, \&
  {Hall}}]{Henry2000ApJS}
{Henry}, G.~W., {Fekel}, F.~C., {Henry}, S.~M., \& {Hall}, D.~S. 2000, \apjs,
  130, 201

\bibitem[{{Hodgkin} {et~al.}(2021){Hodgkin}, {Harrison}, {Breedt}, {Wevers},
  {Rixon}, {Delgado}, {Yoldas}, {Kostrzewa-Rutkowska}, {Wyrzykowski}, {van
  Leeuwen}, {Blagorodnova}, {Campbell}, {Eappachen}, {Fraser}, {Ihanec},
  {Koposov}, {Kruszy{\'n}ska}, {Marton}, {Rybicki}, {Brown}, {Burgess},
  {Busso}, {Cowell}, {De Angeli}, {Diener}, {Evans}, {Gilmore}, {Holland},
  {Jonker}, {van Leeuwen}, {Mignard}, {Osborne}, {Portell}, {Prusti},
  {Richards}, {Riello}, {Seabroke}, {Walton}, {{\'A}brah{\'a}m}, {Altavilla},
  {Baker}, {Bastian}, {O'Brien}, {de Bruijne}, {Butterley}, {Carrasco},
  {Casta{\~n}eda}, {Clark}, {Clementini}, {Copperwheat}, {Cropper},
  {Damljanovic}, {Davidson}, {Davis}, {Dennefeld}, {Dhillon}, {Dolding},
  {Dominik}, {Esquej}, {Eyer}, {Fabricius}, {Fridman}, {Froebrich}, {Garralda},
  {Gomboc}, {Gonz{\'a}lez-Vidal}, {Guerra}, {Hambly}, {Hardy}, {Holl},
  {Hourihane}, {Japelj}, {Kann}, {Kiss}, {Knigge}, {Kolb}, {Komossa},
  {K{\'o}sp{\'a}l}, {Kov{\'a}cs}, {Kun}, {Leto}, {Lewis}, {Littlefair},
  {Mahabal}, {Mundell}, {Nagy}, {Padeletti}, {Palaversa}, {Pigulski},
  {Pretorius}, {van Reeven}, {Ribeiro}, {Roelens}, {Rowell}, {Schartel},
  {Scholz}, {Schwope}, {Sip{\H{o}}cz}, {Smartt}, {Smith}, {Serraller},
  {Steeghs}, {Sullivan}, {Szabados}, {Szegedi-Elek}, {Tisserand}, {Tomasella},
  {van Velzen}, {Whitelock}, {Wilson}, \& {Young}}]{Hodgkin2021}
{Hodgkin}, S.~T., {Harrison}, D.~L., {Breedt}, E., {et~al.} 2021, \aap, 652,
  A76

\bibitem[{{Hodgkin} {et~al.}(2013){Hodgkin}, {Wyrzykowski}, {Blagorodnova}, \&
  {Koposov}}]{Hodgkin2013RSPTA.37120239H}
{Hodgkin}, S.~T., {Wyrzykowski}, L., {Blagorodnova}, N., \& {Koposov}, S. 2013,
  Philosophical Transactions of the Royal Society of London Series A, 371,
  20120239

\bibitem[{{Howell} {et~al.}(2011){Howell}, {Everett}, {Sherry}, {Horch}, \&
  {Ciardi}}]{Howell2011}
{Howell}, S.~B., {Everett}, M.~E., {Sherry}, W., {Horch}, E., \& {Ciardi},
  D.~R. 2011, \aj, 142, 19

\bibitem[{{Ilyin}(2000)}]{2000Ilyin}
{Ilyin}, I.~V. 2000, PhD thesis, University of Oulu, Division of Astronomy

\bibitem[{{Jab{\l}o{\'n}ska} {et~al.}(2022){Jab{\l}o{\'n}ska}, {Wyrzykowski},
  {Rybicki}, {Kruszy{\'n}ska}, {Kaczmarek}, \& {Penoyre}}]{Jablonska2022}
{Jab{\l}o{\'n}ska}, M., {Wyrzykowski}, {\L}., {Rybicki}, K.~A., {et~al.} 2022,
  \aap, 666, L16

\bibitem[{{Jarrett} {et~al.}(2011){Jarrett}, {Cohen}, {Masci}, {Wright},
  {Stern}, {Benford}, {Blain}, {Carey}, {Cutri}, {Eisenhardt}, {Lonsdale},
  {Mainzer}, {Marsh}, {Padgett}, {Petty}, {Ressler}, {Skrutskie}, {Stanford},
  {Surace}, {Tsai}, {Wheelock}, \& {Yan}}]{Jarrett2011ApJ}
{Jarrett}, T.~H., {Cohen}, M., {Masci}, F., {et~al.} 2011, \apj, 735, 112

\bibitem[{{Kaczmarek} {et~al.}(2022){Kaczmarek}, {McGill}, {Evans}, {Smith},
  {Wyrzykowski}, {Howil}, \& {Jab{\l}o{\'n}ska}}]{VVVdarklenses}
{Kaczmarek}, Z., {McGill}, P., {Evans}, N.~W., {et~al.} 2022, \mnras, 514, 4845

\bibitem[{{Kim} {et~al.}(2016){Kim}, {Lee}, {Park}, {Kim}, {Cha}, {Lee}, {Han},
  {Chun}, \& {Yuk}}]{KMTnet2016JKAS...49...37K}
{Kim}, S.-L., {Lee}, C.-U., {Park}, B.-G., {et~al.} 2016, Journal of Korean
  Astronomical Society, 49, 37

\bibitem[{{Kroupa} \& {Weidner}(2003)}]{Kroupa2003}
{Kroupa}, P. \& {Weidner}, C. 2003, \apj, 598, 1076

\bibitem[{{Kruszy{\'n}ska} {et~al.}(2021){Kruszy{\'n}ska}, {Wyrzykowski},
  {Rybicki}, {Maskoli{\={u}}nas}, {Bachelet}, {Rattenbury}, {Mr{\'o}z},
  {Zieli{\'n}ski}, {Howil}, {Kaczmarek}, {Hodgkin}, {Ihanec}, {Gezer},
  {Gromadzki}, {Miko{\l}ajczyk}, {Stankevi{\v{c}}i{\={u}}t{\.{e}}},
  {{\v{C}}epas}, {Pak{\v{s}}tien{\.{e}}}, {{\v{S}}i{\v{s}}kauskait{\.{e}}},
  {Zdanavi{\v{c}}ius}, {Bozza}, {Dominik}, {Figuera Jaimes}, {Fukui},
  {Hundertmark}, {Narita}, {Street}, {Tsapras}, {Bronikowski},
  {Jab{\l}o{\'n}ska}, {Jab{\l}onowska}, \&
  {Zi{\'o}{\l}kowska}}]{Kruszynska2022}
{Kruszy{\'n}ska}, K., {Wyrzykowski}, {\L}., {Rybicki}, K.~A., {et~al.} 2021,
  arXiv e-prints, arXiv:2111.08337

\bibitem[{{Kurucz}(1993)}]{Kurucz1993}
{Kurucz}, R.~L. 1993, {SYNTHE spectrum synthesis programs and line data}

\bibitem[{{Lam} {et~al.}(2022){Lam}, {Lu}, {Udalski}, {Bond}, {Bennett},
  {Skowron}, {Mr{\'o}z}, {Poleski}, {Sumi}, {Szyma{\'n}ski}, {Koz{\l}owski},
  {Pietrukowicz}, {Soszy{\'n}ski}, {Ulaczyk}, {Wyrzykowski}, {Miyazaki},
  {Suzuki}, {Koshimoto}, {Rattenbury}, {Hosek}, {Abe}, {Barry}, {Bhattacharya},
  {Fukui}, {Fujii}, {Hirao}, {Itow}, {Kirikawa}, {Kondo}, {Matsubara},
  {Matsumoto}, {Muraki}, {Olmschenk}, {Ranc}, {Okamura}, {Satoh}, {Silva},
  {Toda}, {Tristram}, {Vandorou}, {Yama}, {Abrams}, {Agarwal}, {Rose}, \&
  {Terry}}]{2022Lam}
{Lam}, C.~Y., {Lu}, J.~R., {Udalski}, A., {et~al.} 2022, \apjl, 933, L23

\bibitem[{{Majewski} {et~al.}(2011){Majewski}, {Zasowski}, \&
  {Nidever}}]{Majewski2011ApJ}
{Majewski}, S.~R., {Zasowski}, G., \& {Nidever}, D.~L. 2011, \apj, 739, 25

\bibitem[{{Marigo} {et~al.}(2013){Marigo}, {Bressan}, {Nanni}, {Girardi}, \&
  {Pumo}}]{Marigo2013}
{Marigo}, P., {Bressan}, A., {Nanni}, A., {Girardi}, L., \& {Pumo}, M.~L. 2013,
  \mnras, 434, 488

\bibitem[{{McCleery} {et~al.}(2020){McCleery}, {Tremblay}, {Gentile Fusillo},
  {Hollands}, {G{\"a}nsicke}, {Izquierdo}, {Toonen}, {Cunningham}, \&
  {Rebassa-Mansergas}}]{McCleery2020MNRAS.499.1890M}
{McCleery}, J., {Tremblay}, P.-E., {Gentile Fusillo}, N.~P., {et~al.} 2020,
  \mnras, 499, 1890

\bibitem[{{Mr{\'o}z} {et~al.}(2022){Mr{\'o}z}, {Udalski}, \&
  {Gould}}]{2022Mroz}
{Mr{\'o}z}, P., {Udalski}, A., \& {Gould}, A. 2022, \apjl, 937, L24

\bibitem[{{Mr{\'o}z} \& {Wyrzykowski}(2021)}]{MrozWyrzykowski2021}
{Mr{\'o}z}, P. \& {Wyrzykowski}, {\L}. 2021, \actaa, 71, 89

\bibitem[{{{\"O}zel} \& {Freire}(2016)}]{NS2016ARA&A..54..401O}
{{\"O}zel}, F. \& {Freire}, P. 2016, \araa, 54, 401

\bibitem[{{{\"O}zel} {et~al.}(2010){{\"O}zel}, {Psaltis}, {Narayan}, \&
  {McClintock}}]{Ozel2010}
{{\"O}zel}, F., {Psaltis}, D., {Narayan}, R., \& {McClintock}, J.~E. 2010,
  \apj, 725, 1918

\bibitem[{{Paczynski}(1986)}]{Paczynski1986}
{Paczynski}, B. 1986, \apj, 304, 1

\bibitem[{{Paczynski}(1996)}]{Paczynski1996}
{Paczynski}, B. 1996, \araa, 34, 419

\bibitem[{{Pecaut} \& {Mamajek}(2013)}]{PecautMamajek2013}
{Pecaut}, M.~J. \& {Mamajek}, E.~E. 2013, \apjs, 208, 9

\bibitem[{{Piascik} {et~al.}(2014){Piascik}, {Steele}, {Bates}, {Mottram},
  {Smith}, {Barnsley}, \& {Bolton}}]{SPRAT2014}
{Piascik}, A.~S., {Steele}, I.~A., {Bates}, S.~D., {et~al.} 2014, in Society of
  Photo-Optical Instrumentation Engineers (SPIE) Conference Series, Vol. 9147,
  Ground-based and Airborne Instrumentation for Astronomy V, ed. S.~K.
  {Ramsay}, I.~S. {McLean}, \& H.~{Takami}, 91478H

\bibitem[{{Poleski} \& {Yee}(2019)}]{PoleskiMulensModel}
{Poleski}, R. \& {Yee}, J.~C. 2019, Astronomy and Computing, 26, 35

\bibitem[{{Raddi} {et~al.}(2022){Raddi}, {Torres}, {Rebassa-Mansergas},
  {Maldonado}, {Camisassa}, {Koester}, {Gentile Fusillo}, {Tremblay}, {Dimpel},
  {Heber}, {Cunningham}, \& {Ren}}]{Raddi2022A&A...658A..22R}
{Raddi}, R., {Torres}, S., {Rebassa-Mansergas}, A., {et~al.} 2022, \aap, 658,
  A22

\bibitem[{{Rybicki} {et~al.}(2022){Rybicki}, {Wyrzykowski}, {Bachelet},
  {Cassan}, {Zieli{\'n}ski}, {Gould}, {Calchi Novati}, {Yee}, {Ryu},
  {Gromadzki}, {Miko{\l}ajczyk}, {Ihanec}, {Kruszy{\'n}ska}, {Hambsch},
  {Zo{\l}a}, {Fossey}, {Awiphan}, {Nakharutai}, {Lewis}, {Olivares E.},
  {Hodgkin}, {Delgado}, {Breedt}, {Harrison}, {van Leeuwen}, {Rixon}, {Wevers},
  {Yoldas}, {Udalski}, {Szyma{\'n}ski}, {Soszy{\'n}ski}, {Pietrukowicz},
  {Koz{\l}owski}, {Skowron}, {Poleski}, {Ulaczyk}, {Mr{\'o}z}, {Iwanek},
  {Wrona}, {Street}, {Tsapras}, {Hundertmark}, {Dominik}, {Beichman}, {Bryden},
  {Carey}, {Gaudi}, {Henderson}, {Shvartzvald}, {Zang}, {Zhu}, {Christie},
  {Green}, {Hennerley}, {McCormick}, {Monard}, {Natusch}, {Pogge}, {Gezer},
  {Gurgul}, {Kaczmarek}, {Konacki}, {Lam}, {Maskoliunas}, {Pakstiene},
  {Ratajczak}, {Stankeviciute}, {Zdanavicius}, \&
  {Zi{\'o}{\l}kowska}}]{RybickiGaia19bld}
{Rybicki}, K.~A., {Wyrzykowski}, {\L}., {Bachelet}, E., {et~al.} 2022, \aap,
  657, A18

\bibitem[{{Rybicki} {et~al.}(2018){Rybicki}, {Wyrzykowski}, {Klencki}, {de
  Bruijne}, {Belczy{\'n}ski}, \& {Chru{\'s}li{\'n}ska}}]{Rybicki2018}
{Rybicki}, K.~A., {Wyrzykowski}, {\L}., {Klencki}, J., {et~al.} 2018, \mnras,
  476, 2013

\bibitem[{{Sahu} {et~al.}(2022){Sahu}, {Anderson}, {Casertano}, {Bond},
  {Udalski}, {Dominik}, {Calamida}, {Bellini}, {Brown}, {Rejkuba}, {Bajaj},
  {Kains}, {Ferguson}, {Fryer}, {Yock}, {Mr{\'o}z}, {Koz{\l}owski},
  {Pietrukowicz}, {Poleski}, {Skowron}, {Soszy{\'n}ski}, {Szyma{\'n}ski},
  {Ulaczyk}, {Wyrzykowski}, {Barry}, {Bennett}, {Bond}, {Hirao}, {Silva},
  {Kondo}, {Koshimoto}, {Ranc}, {Rattenbury}, {Sumi}, {Suzuki}, {Tristram},
  {Vandorou}, {Beaulieu}, {Marquette}, {Cole}, {Fouqu{\'e}}, {Hill}, {Dieters},
  {Coutures}, {Dominis-Prester}, {Bennett}, {Bachelet}, {Menzies}, {Albrow},
  {Pollard}, {Gould}, {Yee}, {Allen}, {Almeida}, {Christie}, {Drummond},
  {Gal-Yam}, {Gorbikov}, {Jablonski}, {Lee}, {Maoz}, {Manulis}, {McCormick},
  {Natusch}, {Pogge}, {Shvartzvald}, {J{\o}rgensen}, {Alsubai}, {Andersen},
  {Bozza}, {Novati}, {Burgdorf}, {Hinse}, {Hundertmark}, {Husser}, {Kerins},
  {Longa-Pe{\~n}a}, {Mancini}, {Penny}, {Rahvar}, {Ricci}, {Sajadian},
  {Skottfelt}, {Snodgrass}, {Southworth}, {Tregloan-Reed}, {Wambsganss},
  {Wertz}, {Tsapras}, {Street}, {Bramich}, {Horne}, {Steele}, \& {RoboNet
  Collaboration}}]{2022Sahu}
{Sahu}, K.~C., {Anderson}, J., {Casertano}, S., {et~al.} 2022, \apj, 933, 83

\bibitem[{{Skowron} {et~al.}(2011){Skowron}, {Udalski}, {Gould}, {Dong},
  {Monard}, {Han}, {Nelson}, {McCormick}, {Moorhouse}, {Thornley}, {Maury},
  {Bramich}, {Greenhill}, {Koz{\l}owski}, {Bond}, {Poleski}, {Wyrzykowski},
  {Ulaczyk}, {Kubiak}, {Szyma{\'n}ski}, {Pietrzy{\'n}ski}, {Soszy{\'n}ski},
  {OGLE Collaboration}, {Gaudi}, {Yee}, {Hung}, {Pogge}, {DePoy}, {Lee},
  {Park}, {Allen}, {Mallia}, {Drummond}, {Bolt}, {{$\mu$}FUN Collaboration},
  {Allan}, {Browne}, {Clay}, {Dominik}, {Fraser}, {Horne}, {Kains}, {Mottram},
  {Snodgrass}, {Steele}, {Street}, {Tsapras}, {RoboNet Collaboration}, {Abe},
  {Bennett}, {Botzler}, {Douchin}, {Freeman}, {Fukui}, {Furusawa}, {Hayashi},
  {Hearnshaw}, {Hosaka}, {Itow}, {Kamiya}, {Kilmartin}, {Korpela}, {Lin},
  {Ling}, {Makita}, {Masuda}, {Matsubara}, {Muraki}, {Nagayama}, {Miyake},
  {Nishimoto}, {Ohnishi}, {Perrott}, {Rattenbury}, {Saito}, {Skuljan},
  {Sullivan}, {Sumi}, {Suzuki}, {Sweatman}, {Tristram}, {Wada}, {Yock}, {MOA
  Collaboration}, {Beaulieu}, {Fouqu{\'e}}, {Albrow}, {Batista}, {Brillant},
  {Caldwell}, {Cassan}, {Cole}, {Cook}, {Coutures}, {Dieters}, {Dominis
  Prester}, {Donatowicz}, {Kane}, {Kubas}, {Marquette}, {Martin}, {Menzies},
  {Sahu}, {Wambsganss}, {Williams}, {Zub}, \& {PLANET
  Collaboration}}]{Skowron2011}
{Skowron}, J., {Udalski}, A., {Gould}, A., {et~al.} 2011, \apj, 738, 87

\bibitem[{{Skrutskie} {et~al.}(2006){Skrutskie}, {Cutri}, {Stiening},
  {Weinberg}, {Schneider}, {Carpenter}, {Beichman}, {Capps}, {Chester},
  {Elias}, {Huchra}, {Liebert}, {Lonsdale}, {Monet}, {Price}, {Seitzer},
  {Jarrett}, {Kirkpatrick}, {Gizis}, {Howard}, {Evans}, {Fowler}, {Fullmer},
  {Hurt}, {Light}, {Kopan}, {Marsh}, {McCallon}, {Tam}, {Van Dyk}, \&
  {Wheelock}}]{Skrutskie:2006cat}
{Skrutskie}, M.~F., {Cutri}, R.~M., {Stiening}, R., {et~al.} 2006, \aj, 131,
  1163

\bibitem[{{Smith} {et~al.}(2002){Smith}, {Mao}, {Wo{\'z}niak}, {Udalski},
  {Szyma{\'n}ski}, {Kubiak}, {Pietrzy{\'n}ski}, {Soszy{\'n}ski}, \&
  {{\.Z}ebru{\'n}}}]{Smith2002}
{Smith}, M.~C., {Mao}, S., {Wo{\'z}niak}, P., {et~al.} 2002, \mnras, 336, 670

\bibitem[{{Stassun} {et~al.}(2019){Stassun}, {Oelkers}, {Paegert}, {Torres},
  {Pepper}, {De Lee}, {Collins}, {Latham}, {Muirhead}, {Chittidi},
  {Rojas-Ayala}, {Fleming}, {Rose}, {Tenenbaum}, {Ting}, {Kane}, {Barclay},
  {Bean}, {Brassuer}, {Charbonneau}, {Ge}, {Lissauer}, {Mann}, {McLean},
  {Mullally}, {Narita}, {Plavchan}, {Ricker}, {Sasselov}, {Seager}, {Sharma},
  {Shiao}, {Sozzetti}, {Stello}, {Vanderspek}, {Wallace}, \&
  {Winn}}]{StassunTESS}
{Stassun}, K.~G., {Oelkers}, R.~J., {Paegert}, M., {et~al.} 2019, \aj, 158, 138

\bibitem[{{Steele} {et~al.}(2004){Steele}, {Smith}, {Rees}, {Baker}, {Bates},
  {Bode}, {Bowman}, {Carter}, {Etherton}, {Ford}, {Fraser}, {Gomboc}, {Lett},
  {Mansfield}, {Marchant}, {Medrano-Cerda}, {Mottram}, {Raback}, {Scott},
  {Tomlinson}, \& {Zamanov}}]{2004SteeleLT}
{Steele}, I.~A., {Smith}, R.~J., {Rees}, P.~C., {et~al.} 2004, in Society of
  Photo-Optical Instrumentation Engineers (SPIE) Conference Series, Vol. 5489,
  Ground-based Telescopes, ed. J.~{Oschmann}, Jacobus~M., 679--692

\bibitem[{{Strai{\v{z}}ys}(1992)}]{Straizys1992}
{Strai{\v{z}}ys}, V. 1992, {Multicolor stellar photometry}

\bibitem[{{Strai{\v{z}}ys} \& {Lazauskait{\.{e}}}(2009)}]{StraizysLazauskaite}
{Strai{\v{z}}ys}, V. \& {Lazauskait{\.{e}}}, R. 2009, Baltic Astronomy, 18, 19

\bibitem[{{Strassmeier} {et~al.}(2015){Strassmeier}, {Ilyin}, {J{\"a}rvinen},
  {Weber}, {Woche}, {Barnes}, {Bauer}, {Beckert}, {Bittner}, {Bredthauer},
  {Carroll}, {Denker}, {Dionies}, {DiVarano}, {D{\"o}scher}, {Fechner},
  {Feuerstein}, {Granzer}, {Hahn}, {Harnisch}, {Hofmann}, {Lesser}, {Paschke},
  {Pankratow}, {Plank}, {Pl{\"u}schke}, {Popow}, \& {Sablowski}}]{strass2015}
{Strassmeier}, K.~G., {Ilyin}, I., {J{\"a}rvinen}, A., {et~al.} 2015,
  Astronomische Nachrichten, 336, 324

\bibitem[{{Sumi} {et~al.}(2013){Sumi}, {Bennett}, {Bond}, {Abe}, {Botzler},
  {Fukui}, {Furusawa}, {Itow}, {Ling}, {Masuda}, {Matsubara}, {Muraki},
  {Ohnishi}, {Rattenbury}, {Saito}, {Sullivan}, {Suzuki}, {Sweatman},
  {Tristram}, {Wada}, {Yock}, \& {MOA Collaboratoin}}]{MOA2013ApJ...778..150S}
{Sumi}, T., {Bennett}, D.~P., {Bond}, I.~A., {et~al.} 2013, \apj, 778, 150

\bibitem[{{Takahashi} {et~al.}(2013){Takahashi}, {Yoshida}, \&
  {Umeda}}]{Takahashi2013ApJ...771...28T}
{Takahashi}, K., {Yoshida}, T., \& {Umeda}, H. 2013, \apj, 771, 28

\bibitem[{{Udalski} {et~al.}(2015){Udalski}, {Szyma{\'n}ski}, \&
  {Szyma{\'n}ski}}]{Udalski2015}
{Udalski}, A., {Szyma{\'n}ski}, M.~K., \& {Szyma{\'n}ski}, G. 2015, \actaa, 65,
  1

\bibitem[{{Veltz} {et~al.}(2008){Veltz}, {Bienaym{\'e}}, {Freeman}, {Binney},
  {Bland-Hawthorn}, {Gibson}, {Gilmore}, {Grebel}, {Helmi}, {Munari},
  {Navarro}, {Parker}, {Seabroke}, {Siebert}, {Steinmetz}, {Watson},
  {Williams}, {Wyse}, \& {Zwitter}}]{Veltz}
{Veltz}, L., {Bienaym{\'e}}, O., {Freeman}, K.~C., {et~al.} 2008, \aap, 480,
  753

\bibitem[{{Volgenau} {et~al.}(2022){Volgenau}, {Harbeck}, {Lindstrom},
  {Collom}, {Street}, \& {Johnson}}]{VolgenauTOM2022}
{Volgenau}, N., {Harbeck}, D., {Lindstrom}, W., {et~al.} 2022, in Society of
  Photo-Optical Instrumentation Engineers (SPIE) Conference Series, Vol. 12186,
  Observatory Operations: Strategies, Processes, and Systems IX, ed. D.~S.
  {Adler}, R.~L. {Seaman}, \& C.~R. {Benn}, 121860W

\bibitem[{{Wang} \& {Chen}(2019)}]{Wang2019}
{Wang}, S. \& {Chen}, X. 2019, \apj, 877, 116

\bibitem[{{Wright} {et~al.}(2010){Wright}, {Eisenhardt}, {Mainzer}, {Ressler},
  {Cutri}, {Jarrett}, {Kirkpatrick}, {Padgett}, {McMillan}, {Skrutskie},
  {Stanford}, {Cohen}, {Walker}, {Mather}, {Leisawitz}, {Gautier}, {McLean},
  {Benford}, {Lonsdale}, {Blain}, {Mendez}, {Irace}, {Duval}, {Liu}, {Royer},
  {Heinrichsen}, {Howard}, {Shannon}, {Kendall}, {Walsh}, {Larsen}, {Cardon},
  {Schick}, {Schwalm}, {Abid}, {Fabinsky}, {Naes}, \& {Tsai}}]{Wright2010AJ}
{Wright}, E.~L., {Eisenhardt}, P. R.~M., {Mainzer}, A.~K., {et~al.} 2010, \aj,
  140, 1868

\bibitem[{{Wyrzykowski} \& {Hodgkin}(2012)}]{Wyrzykowski2012}
{Wyrzykowski}, {\L}. \& {Hodgkin}, S. 2012, in IAU Symposium, Vol. 285, IAU
  Symposium, ed. E.~{Griffin}, R.~{Hanisch}, \& R.~{Seaman}, 425--428

\bibitem[{{Wyrzykowski} {et~al.}(2016){Wyrzykowski}, {Kostrzewa-Rutkowska},
  {Skowron}, {Rybicki}, {Mr{\'o}z}, {Koz{\l}owski}, {Udalski}, {Szyma{\'n}ski},
  {Pietrzy{\'n}ski}, {Soszy{\'n}ski}, {Ulaczyk}, {Pietrukowicz}, {Poleski},
  {Pawlak}, {I{\l}kiewicz}, \& {Rattenbury}}]{Wyrz16}
{Wyrzykowski}, {\L}., {Kostrzewa-Rutkowska}, Z., {Skowron}, J., {et~al.} 2016,
  \mnras, 458, 3012

\bibitem[{{Wyrzykowski} {et~al.}(2009){Wyrzykowski}, {Koz{\l}owski}, {Skowron},
  {Belokurov}, {Smith}, {Udalski}, {Szyma{\'n}ski}, {Kubiak},
  {Pietrzy{\'n}ski}, {Soszy{\'n}ski}, {Szewczyk}, \&
  {{\.Z}ebru{\'n}}}]{Wyrzykowski2009}
{Wyrzykowski}, {\L}., {Koz{\l}owski}, S., {Skowron}, J., {et~al.} 2009, \mnras,
  397, 1228

\bibitem[{{Wyrzykowski} {et~al.}(2023){Wyrzykowski}, {Kruszy{\'n}ska},
  {Rybicki}, {Holl}, {Lec{\oe}ur-Ta{\"\i}bi}, {Mowlavi}, {Nienartowicz},
  {Jevardat de Fombelle}, {Rimoldini}, {Audard}, {Garcia-Lario}, {Gavras},
  {Evans}, {Hodgkin}, \& {Eyer}}]{Wyrzykowski2023}
{Wyrzykowski}, {\L}., {Kruszy{\'n}ska}, K., {Rybicki}, K.~A., {et~al.} 2023,
  \aap, 674, A23

\bibitem[{{Wyrzykowski} \& {Mandel}(2020)}]{WyrzykowskiMandel2020}
{Wyrzykowski}, {\L}. \& {Mandel}, I. 2020, \aap, 636, A20

\bibitem[{{Wyrzykowski} {et~al.}(2020){Wyrzykowski}, {Mr{\'o}z}, {Rybicki},
  {Gromadzki}, {Ko{\l}aczkowski}, {Zieli{\'n}ski}, {Zieli{\'n}ski},
  {Britavskiy}, {Gomboc}, {Sokolovsky}, {Hodgkin}, {Abe}, {Aldi}, {AlMannaei},
  {Altavilla}, {Al Qasim}, {Anupama}, {Awiphan}, {Bachelet}, {Bak{\i}{\c{s}}},
  {Baker}, {Bartlett}, {Bendjoya}, {Benson}, {Bikmaev}, {Birenbaum},
  {Blagorodnova}, {Blanco-Cuaresma}, {Boeva}, {Bonanos}, {Bozza}, {Bramich},
  {Bruni}, {Burenin}, {Burgaz}, {Butterley}, {Caines}, {Caton}, {Calchi
  Novati}, {Carrasco}, {Cassan}, {{\v{C}}epas}, {Cropper},
  {Chru{\'s}li{\'n}ska}, {Clementini}, {Clerici}, {Conti}, {Conti}, {Cross},
  {Cusano}, {Damljanovic}, {Dapergolas}, {D'Ago}, {de Bruijne}, {Dennefeld},
  {Dhillon}, {Dominik}, {Dziedzic}, {Erece}, {Eselevich}, {Esenoglu}, {Eyer},
  {Figuera Jaimes}, {Fossey}, {Galeev}, {Grebenev}, {Gupta}, {Gutaev},
  {Hallakoun}, {Hamanowicz}, {Han}, {Handzlik}, {Haislip}, {Hanlon}, {Hardy},
  {Harrison}, {van Heerden}, {Hoette}, {Horne}, {Hudec}, {Hundertmark},
  {Ihanec}, {Irtuganov}, {Itoh}, {Iwanek}, {Jovanovic}, {Janulis},
  {Jel{\'\i}nek}, {Jensen}, {Kaczmarek}, {Katz}, {Khamitov}, {Kilic},
  {Klencki}, {Kolb}, {Kopacki}, {Kouprianov}, {Kruszy{\'n}ska}, {Kurowski},
  {Latev}, {Lee}, {Leonini}, {Leto}, {Lewis}, {Li}, {Liakos}, {Littlefair},
  {Lu}, {Manser}, {Mao}, {Maoz}, {Martin-Carrillo}, {Marais},
  {Maskoli{\={u}}nas}, {Maund}, {Meintjes}, {Melnikov}, {Ment},
  {Miko{\l}ajczyk}, {Morrell}, {Mowlavi}, {Mo{\'z}dzierski}, {Murphy},
  {Nazarov}, {Netzel}, {Nesci}, {Ngeow}, {Norton}, {Ofek},
  {Pak{\v{s}}tien{\.{e}}}, {Palaversa}, {Pandey}, {Paraskeva}, {Pawlak},
  {Penny}, {Penprase}, {Piascik}, {Prieto}, {Qvam}, {Ranc},
  {Rebassa-Mansergas}, {Reichart}, {Reig}, {Rhodes}, {Rivet}, {Rixon},
  {Roberts}, {Rosi}, {Russell}, {Zanmar Sanchez}, {Scarpetta}, {Seabroke},
  {Shappee}, {Schmidt}, {Shvartzvald}, {Sitek}, {Skowron}, {{\'S}niegowska},
  {Snodgrass}, {Soares}, {van Soelen}, {Spetsieri},
  {Stankevi{\v{c}}i{\={u}}t{\.{e}}}, {Steele}, {Street}, {Strobl}, {Strubble},
  {Szegedi}, {Tinjaca Ramirez}, {Tomasella}, {Tsapras}, {Vernet}, {Villanueva},
  {Vince}, {Wambsganss}, {van der Westhuizen}, {Wiersema}, {Wium}, {Wilson},
  {Yoldas}, {Zhuchkov}, {Zhukov}, {Zdanavi{\v{c}}ius}, {Zo{\l}a}, \&
  {Zubareva}}]{WyrzykowskiGaia16aye}
{Wyrzykowski}, {\L}., {Mr{\'o}z}, P., {Rybicki}, K.~A., {et~al.} 2020, \aap,
  633, A98

\bibitem[{{Wyrzykowski} {et~al.}(2011){Wyrzykowski}, {Skowron}, {Koz{\l}owski},
  {Udalski}, {Szyma{\'n}ski}, {Kubiak}, {Pietrzy{\'n}ski}, {Soszy{\'n}ski},
  {Szewczyk}, {Ulaczyk}, {Poleski}, \& {Tisserand}}]{Wyrzykowski2011b}
{Wyrzykowski}, L., {Skowron}, J., {Koz{\l}owski}, S., {et~al.} 2011, \mnras,
  416, 2949

\bibitem[{{Zieli{\'n}ski} {et~al.}(2020){Zieli{\'n}ski}, {Wyrzykowski},
  {Miko{\l}ajczyk}, {Rybicki}, \& {Ko{\l}aczkowski}}]{2020CPCS2}
{Zieli{\'n}ski}, P., {Wyrzykowski}, {\l}., {Miko{\l}ajczyk}, P., {Rybicki}, K.,
  \& {Ko{\l}aczkowski}, Z. 2020, in XXXIX Polish Astronomical Society Meeting,
  ed. K.~{Ma{\l}ek}, M.~{Poli{\'n}ska}, A.~{Majczyna}, G.~{Stachowski},
  R.~{Poleski}, {\L}.~{Wyrzykowski}, \& A.~{R{\'o}{\.z}a{\'n}ska}, Vol.~10,
  190--193

\bibitem[{{Zieli{\'n}ski} {et~al.}(2019){Zieli{\'n}ski}, {Wyrzykowski},
  {Rybicki}, {Ko{\l}aczkowski}, {Bru{\'s}}, \& {Miko{\l}ajczyk}}]{2019CPCS2}
{Zieli{\'n}ski}, P., {Wyrzykowski}, {\L}., {Rybicki}, K., {et~al.} 2019,
  Contributions of the Astronomical Observatory Skalnate Pleso, 49, 125

\end{thebibliography}
\bibliographystyle{aa}

\section{Appendix}
\subsection{Photometry}
Here we present the photometric data used for modelling Gaia19dke.

\begin{table*} 
\begin{scriptsize}
\centering
\caption{Telescopes involved in the photometric follow-up observations of Gaia19dke}
\footnotesize
\label{tab:observatories}
\begin{tabular}{llllll}
\hline
Observatory & Name & Location & Longitude & Latitude & Ref.  \\
 &  &  & [deg] (E+) & [deg] &   \\
\hline
\hline
ASV & 1.4-m Milankovic Telescope & Vidojevica, Serbia & 21.56 & 43.14 & 1 \\ 
 & 60-cm Nedeljkovic Telescope & Vidojevica, Serbia & 21.56 & 43.14 & 1\\
 & Astronomical Observatory of Belgrade &  &  &  &  \\
Abastumani & 36-cm telescope & Mount Kanobili,  & 42.82 & 41.75 & 2 \\
 & Georgian National Astrophysical Observatory & Georgia &  &  &  \\
Adiyaman 60 & 60-cm telescope & Adiyaman, Turkey & 41.23 & 39.78 & 3 \\
& Adiyaman University Observatory & & & & \\
Adonis & 25-cm Adonis Observatory telescope & Langemark, Belgium  & 2.93 & 50.92 & - \\
Astrolab & 68-cm NMPT telescope & Ypres, Belgium & 2.91 & 50.82 & 4 \\
& Astrolab IRIS Observatory & & & & \\
Bia\l{}k\'ow & 60-cm Cassegrain telescope, Bia\l{}k\'ow Observatory  & Bia\l{}k\'ow, Poland & 16.66 & 51.48 & 5 \\ 
 & Astronomical Institute, University of Wroc\l{}aw &  &  &  &  \\
 Flarestar & 25-cm Schmidt-Cassegrain telescope & San Gwann,  & 14.47 & 35.90 & 6 \\
 & Flarestar Observatory & Malta & & & \\
 HAO & 68-cm Horten telescope & Nykirke, Norway & 10.39 & 59.43 & - \\
 IST60 & 60-cm Ritchey-Chretien telescope & Ulupinar, Turkey & 26.47 & 40.10 & 7 \\
 Krakow CDK500 & 50-cm telescope & Krak\'ow, Poland & 19.83 & 50.05 & 8 \\
 & Astronomical Observatory of the Jagiellonian University & & & & \\
 LCO 1m & 1.0-m telescope & Texas, US & -104.02 & 30.67 & 9 \\
 & McDonald Observatory & & & & \\
 LCO 1m & 1.0-m telescope & Izaña, Tenerife & -16.51 & 28.30 & 9 \\
 & Tenerife Observatory & Spain & & & \\
 Loiano & 1.52-m Cassini Telescope, & Bologna, Loiano, Italy & 11.33 & 44.26 & 10 \\ 
 & Bologna Observatory of Astrophysics and Space Science & & &  &  \\
 Moletai35 & 35-cm Maksutov telescope,  & Mol\.etai, Kulionys, & 25.56 & 55.32 & 11 \\
 &  Mol\.etai Astronomical Observatory & Lithuania &  &  &  \\
Ostrowik60 & 60-cm Cassegrain telescope,  & Ostrowik, Poland & 21.42 & 52.09 & 12\\
 &  Warsaw University Astronomical Observatory &  &  &  & \\
 Piwnice 90 & 90-cm Schmidt-Cassegrain telescope & Piwnice, Poland & 18.56 & 53.09 & 13 \\
 & Institute of Astronomy, Nicolaus Copernicus University & & & & \\
 RBT & 70-cm CDK telescope, Adam Mickiewicz University & Winer Observatory, AZ, USA & -110.60 & 31.66 & 24 \\
 Rozhen60 & 60-cm Cassegrain telescope & Rozhen, Bulgaria & 24.74 & 41.7 & 14 \\
 & Rozhen National Astronomical Observatory & & & & \\
 RRRT & 60-cm Rapid Response Robotic Telescope & Virginia, US & -78.69 & 37.88 & 15 \\
 & Fan Mountain Observatory & & & & \\
 SUTO-Otivar & 30-cm Newtonian telescope & Otivar, Spain & -3.68 & 36.82 & 16 \\
 & Silesian University of Technology Observatory &  &  &  & \\
 SUTO-Pyskowice & 30-cm Newtonian telescope & Pyskowice, Poland & 18.63 & 50.39  & 16 \\
 & Silesian University of Technology Observatory &  &  &  & \\
T100 & 1.0-m Ritchey–Chrétien telescope & Bakırlıtepe, Turkey & 30.33 & 36.82 & 17 \\
 & TÜBİTAK National Observatory & & & & \\
 TJO & 80-cm Joan Or\'o Telescope, Montsec Observatory & Sant Esteve de la Sarga & 0.73 & 42.03 & 18 \\
  & Observatori Astron\'omic del Montsec  & Lleida, Spain &  &  &  \\
TRT & Thai Robotic Telescope GAO, Yunnan Observatory & Phoenix Mountain & 105.03 & 26.70 & 19 \\
 &  & Kunming, China & &  &  \\
 Terskol2m & 2.0-m Zeiss Ritchey-Chretien-Coude telescope & North Caucasus & 43.27 & 42.5 & - \\
 & Terskol Ukrainian Observatory & & & & \\
 Tomo-e Gozen &  1.05-m Schmidt telescope & Kiso, Nagano, Japan & 137.63 & 35.80 & 20 \\
 & Kiso Observatory, the University of Tokyo & & & & \\
UZPW50 & 50-cm telescope & e-EyE, Spain & -6.63 & 38.22 & 21 \\
& University of Zielona G\'ora & & & & \\
VATT & Vatican Advanced Technology Telescope & Mount Graham,  & -109.72 & 32.72 & 22 \\
 &  & Arizona, US  &  &  &  \\
ZAO & 20-cm SCT Telescope, & Malta & 14.39 & 35.85 & 23 \\
& Znith Astronomy Observatory & & & & \\
\hline
\hline
\end{tabular}
\end{scriptsize}
\begin{tiny}
References: 1: \url{http://vidojevica.aob.rs/}, \cite{ASV}, 2: \url{http://www.abao.ge/en/}, 3: \url{https://observatory.adiyaman.edu.tr/}, 4: \url{https://astrolab.be/}, 5: \url{https://uwr.edu.pl/en/visit-us/bialkovo-observatory/}, 6: \url{https://flarestar.weebly.com/}, 7: \url{https://caam.comu.edu.tr/}, 8: \url{http://www.oa.uj.edu.pl/}, 9: \url{https://lco.global/observatory/}, \cite{LCO} 10: \url{https://www.oas.inaf.it/en/}, 11: \url{http://mao.tfai.vu.lt/sci/en/news/}, 12: \url{https://ostrowik.astrouw.edu.pl/}, 13: \url{https://astro.umk.pl/en/}, 14: \url{https://www.rozhen.org/}, 15: \url{https://astronomy.as.virginia.edu/research/observatories/fan-mountain}, 16: \url{https://www.suto.aei.polsl.pl/}, 17: \url{https://tug.tubitak.gov.tr/en}, 18: \url{https://www.ieec.cat/content/206/what-s-the-oadm/}, 19: \url{https://trt.narit.or.th/obsinfo/gao}, 20: \url{http://www.ioa.s.u-tokyo.ac.jp/kisohp/top_e.html}, 21: \url{https://ia.uz.zgora.pl/}, 22: \url{https://www.vaticanobservatory.org/}, 23: \url{https://znith-observatory.blogspot.com/}, 24: \url{http://pallas.astro.amu.edu.pl/~chrisk/gats/}

\end{tiny}
\end{table*}


\begin{table*} 
\begin{scriptsize}
\centering
\caption{Photometric data collected for Gaia19dke with a network of follow-up telescopes}
\footnotesize
\label{tab:photometrystats}
\begin{tabular}{lllll}
\hline
Observatory  &  Filters$^*$ & $N_{points}$ & First MJD & Last MJD  \\
\hline
\hline
ASV & B, I, R, U, V, g, r, z  & 160         & 58724.9 & 59518.7 \\
Abastumani & B, I, R, V, i, r    & 46         & 59034.9 & 59870.8 \\
Adiyaman60 & R, g, i, r, u     & 41         & 59031.8 & 59342.9 \\
Adonis & B, I, R, V, r     & 159         & 59760.9 & 59946.7 \\
Astrolab & B, I, R, V      & 37         & 58760.9 & 59515.8 \\
Bialkow & B, I, R, V      & 20         & 59649.1 & 59649.2 \\
Flarestar & I, V        & 17         & 59378.9 & 59442.8 \\
HAO & B, I, R, V, r     & 37         & 59315.1 & 59691.0 \\
IST60 & R, V, i       & 4         & 58761.8 & 58762.9 \\
KrakowCDK500 & I, R, V       & 46         & 59072.9 & 59859.8 \\
LCO1m & g, i, r       & 277         & 58807.0 & 60097.0 \\
Loiano & I, R        & 3         & 59046.9 & 59046.9 \\
Moletai35 & I, R, V, r      & 293         & 58750.0 & 59715.9 \\
Ostrowik60 & I, R, i, u, z     & 26         & 59086.8 & 59447.9 \\
Piwnice90 & B, I, R, U, V, i, r, z  & 104         & 59649.1 & 59828.8 \\
RBT & U, B, V, R, u, g, r & 213 & 59151.6 & 59496.1 \\
RRRT & R, V        & 2         & 59840.2 & 59840.2 \\
SUTO-Otivar & B, I, V, i, r     & 309         & 59708.2 & 60016.2 \\
SUTO-Pyskowice & B, I, R, V, i, r    & 60         & 59700.1 & 59749.9 \\
T100 & g, i        & 3         & 59149.7 & 59149.7 \\
TJO & B, I, R, U, V, g, r, u, z & 1176 & 58730.8 & 60025.1 \\
TRT & I, V        & 7         & 58914.9 & 58939.8 \\
Terskol2m & B, I, R       & 11         & 59431.0 & 59436.5 \\
Tomo-e Gozen & R, r, u       & 18 & 58757.4 & 59066.5 \\
UZPW50 & g, i, r       & 66         & 59711.1 & 59811.9 \\
ZAO & I         & 7         & 59619.2 & 59620.2 \\
\hline
Gaia & G & 191 & 56960.1 & 60061.4\\
ZTF & g,r & 1491 & 58203.5 & 60065.4\\ 
\hline
\hline
\end{tabular}

$^*$ List of filters the original observations were standardised to using Gaia Synthetic Photometry catalogue. Capital letters denote Johnson-Kron-Cousins bands, while lowercase letters denote SDSS bands.
\end{scriptsize}
\end{table*}

\begin{table}
      \caption{Gaia19dke photometry from {\gaia} Science Alerts during 2014-10-30 to 2022-10-09 time period. The uncertainties (err[mag]) were estimated using {\gaia} DR2. The full table is available at the CDS.}

         \label{tab:photGaia}
          \begin{tabular}{|c|c|c|c|} 
          \hline
            \noalign{\smallskip}
           Date & JD &  G[mag] & err[mag] \\
            \noalign{\smallskip}
            \hline
            \noalign{\smallskip}
    2014-10-30 02:51:46  & 2456960.61928  & 15.53  & 0.0081 \\
   2014-10-30 04:38:20  & 2456960.69329  & 15.51  & 0.0080 \\
          ...              &   ...          & ...    & ...    \\
      2018-05-18 08:30:17  & 2458256.85436  & 15.46  & 0.0078 \\
   2018-06-30 11:33:01  & 2458299.98126  & 15.45  & 0.0078 \\
      ...              &   ...          & ...    & ...    \\
   2019-08-08 02:58:44  & 2458703.62412  & 15.23  & 0.0070 \\
   2019-08-08 04:45:18  & 2458703.69813  & 15.22  & 0.0070 \\
   2019-08-08 08:58:58  & 2458703.87428  & 15.22  & 0.0070 \\
   2019-08-08 10:45:32  & 2458703.94829  & 15.24  & 0.0070 \\
   2019-08-28 17:12:34  & 2458724.21706  & 15.19  & 0.0069 \\
   2019-08-28 21:26:13  & 2458724.39321  & 15.21  & 0.0070 \\
   2019-08-28 23:12:47  & 2458724.46721  & 15.20  & 0.0069 \\
   2019-11-10 06:36:36  & 2458797.77542  & 15.15  & 0.0068 \\
   2019-11-10 08:23:10  & 2458797.84942  & 15.12  & 0.0067 \\
   2019-12-01 01:17:13  & 2458818.55362  & 15.18  & 0.0069 \\
   2019-12-01 03:03:46  & 2458818.62762  & 15.14  & 0.0067 \\
   2020-01-12 05:46:44  & 2458860.74079  & 15.23  & 0.0070 \\
   2020-01-12 10:00:24  & 2458860.91694  & 15.20  & 0.0069 \\
   2020-01-12 11:46:58  & 2458860.99095  & 15.19  & 0.0069 \\
   2020-01-20 16:10:04  & 2458869.17366  & 15.24  & 0.0070 \\
   2020-01-20 17:56:38  & 2458869.24766  & 15.21  & 0.0070 \\
   2020-03-04 01:03:47  & 2458912.54429  & 15.22  & 0.0070 \\
   2020-03-04 02:50:21  & 2458912.61830  & 15.21  & 0.0070 \\
   2020-04-27 09:31:30  & 2458966.89688  & 15.13  & 0.0067 \\
   2020-06-03 06:54:09  & 2459003.78760  & 15.03  & 0.0064 \\
   2020-06-26 17:45:16  & 2459027.23977  & 14.93  & 0.0061 \\
   2020-08-26 16:03:52  & 2459088.16935  & 14.87  & 0.0059 \\
   2020-08-26 17:50:26  & 2459088.24336  & 14.86  & 0.0059 \\
   2020-09-11 22:07:22  & 2459104.42178  & 14.91  & 0.0061 \\
   2020-09-11 23:53:56  & 2459104.49579  & 14.92  & 0.0061 \\
   2020-10-19 15:12:20  & 2459142.13356  & 15.07  & 0.0065 \\
   2020-10-19 16:58:54  & 2459142.20757  & 15.08  & 0.0065 \\
   2020-11-20 02:48:28  & 2459173.61699  & 15.15  & 0.0068 \\
   2020-12-15 02:46:53  & 2459198.61589  & 15.20  & 0.0069 \\
   2020-12-15 04:33:27  & 2459198.68990  & 15.20  & 0.0069 \\
   2021-02-04 04:14:05  & 2459249.67645  & 15.21  & 0.0070 \\
   2021-02-04 08:27:46  & 2459249.85262  & 15.24  & 0.0070 \\
   2021-02-04 10:14:20  & 2459249.92662  & 15.21  & 0.0070 \\
   2021-02-23 20:38:16  & 2459269.35991  & 15.20  & 0.0069 \\
   2021-02-23 22:24:50  & 2459269.43391  & 15.21  & 0.0070 \\
   2021-02-24 02:38:31  & 2459269.61008  & 15.20  & 0.0069 \\
   2021-02-24 04:25:06  & 2459269.68410  & 15.21  & 0.0070 \\
   2021-04-04 02:16:04  & 2459308.59449  & 15.19  & 0.0069 \\
   2021-04-04 04:02:38  & 2459308.66850  & 15.18  & 0.0069 \\
   2021-05-08 07:46:52  & 2459342.82421  & 15.16  & 0.0068 \\
   2021-05-08 09:33:27  & 2459342.89823  & 15.15  & 0.0068 \\
   2021-05-30 01:46:48  & 2459364.57417  & 15.16  & 0.0068 \\
   2021-07-14 00:26:14  & 2459409.51822  & 15.21  & 0.0070 \\
   2021-07-14 02:12:48  & 2459409.59222  & 15.21  & 0.0070 \\
   2021-07-14 06:26:28  & 2459409.76838  & 15.20  & 0.0069 \\
   2021-07-14 08:13:03  & 2459409.84240  & 15.22  & 0.0070 \\
    ...                 &  ...           & ...    & ...    \\
   2022-08-04 03:14:13  & 2459795.63487  & 15.44  & 0.0077 \\
   2022-08-04 07:27:53  & 2459795.81103  & 15.44  & 0.0077 \\
   2022-09-17 12:05:05  & 2459840.00353  & 15.45  & 0.0078 \\
   2022-10-09 13:45:50  & 2459862.07350  & 15.47  & 0.0078 \\

            \noalign{\smallskip}
            \hline
         \end{tabular}
\end{table}

\end{document}